\documentclass[aps,prl,
twocolumn,superscriptaddress,floatfix,fleqn]{revtex4}
\usepackage{graphicx}
\usepackage{bm}
\usepackage{amsmath,amssymb}
\renewcommand{\Im}{\mathop{\rm Im}}
\renewcommand{\Re}{\mathop{\rm Re}}
\begin{document}
\title{Resonant nonlinear optics of backward waves in negative-index
metamaterials
}

\author{Alexander K. Popov}
\affiliation{Department of Physics \& Astronomy, University of Wisconsin-Stevens Point,
Stevens Point, WI 54481, USA}
\email{apopov@uwsp.edu}
\author{Sergei A. Myslivets}
\affiliation{Institute of Physics of the Russian Academy of Sciences, 660036 Krasnoyarsk,
Russian Federation}
\author{Vladimir M.
Shalaev}
\affiliation{Birck Nanotechnology Center and School of Electrical and Computer
Engineering, Purdue University, West Lafayette, IN 47907, USA}

\begin{abstract}The extraordinary properties of resonant four-wave mixing of backward waves in doped negative-index materials are investigated.  The feasibility of independent engineering of negative refractive index and nonlinear optical response as well as quantum control of the nonlinear propagation process in such composites is shown due to the coherent energy transfer from the control to the signal field.  Laser-induced transparency, quantum switching, frequency-tunable narrow-band filtering, amplification, and realizing a miniature mirrorless optical parametric generator of the entangled backward and ordinary waves are among the possible applications of the investigated processes.   
\end{abstract}
\maketitle



\section{Introduction}
Optical negative index metamaterials (NIMs) with simultaneously negative electric and magnetic responses (NIMs) promise revolutionary breakthroughs in photonics (see, for example, \cite{Sh}). The most detrimental obstacle toward the application of NIMs is strong the absorption of light that is inherent to this class of materials. The possibility to overcome such obstacles based on three-wave [$\chi ^{(2)}$] optical parametric amplification (OPA) in NIMs was proposed in \cite{APB,OL}. However, a strong nonlinear-optical (NLO) response must be ensured and the corresponding frequency domain must overlap the phase-matching frequency-interval to realize such feasibility. Nonlinear optics in NIMs remains so far a less-developed branch of optics. The possibility of quadratic NLO optical response in NIMs attributed to the asymmetry of the voltage-current characteristics of their nanoscaled building blocks was predicted in \cite {Lap,Kiv}. Recent experimental demonstration of the exciting feasibilities to craft NIMs with strong NLO response has been reported in \cite{Kl}. Counterintuitive properties of nonlinear propagation processes in NIMs with $ \chi ^{(2)}$ NLO response, such as second harmonic generation and three-wave OPA, as compared with their counterparts in natural materials, were revealed in \cite{APB,OL,Agr,SHG,Sc}. The striking changes in the optical bistability in layered structures including MIM were shown in \cite{Lit}. A review of the corresponding theoretical approaches is given in \cite{Gab}. Extraordinary properties of a three-wave mixing (TWM) backward-wave (BW) optical parametric oscillator (OPO) were predicted several decades ago in \cite{Har,Vol,Yar}. However, phase matching of the coupled waves with the \emph{opposite} orientation of their wave vectors, which is required for mirrorless oscillations in ordinary materials, is very difficult to achieve for the waves with substantially different frequencies. For the first time, TWM backward-wave mirrorless optical parametrical oscillator (BWMOPO) with all three different optical wavelengths was realized only recently \cite {Kh,Pas}. Phase matching of waves with an antiparallel orientation of their wave vectors has been achieved in a submicrometer periodically poled NLO crystal, which has become possible owing to recent advances in nanotechnology. As outlined, a major technical problem in creating BWMOPO stems from the requirement of phase matching for traveling waves with oppositely oriented wave vectors, a situation that is inherent to ordinary materials. However, the situation dramatically changes in NIMs. Due to the opposite orientation of the energy flow and wave vector, which is inherent to NIMs, phase matching of backward waves becomes possible for the \emph{ parallel} orientation of wave-vectors for all coupled waves. The extraordinary, distributed-feedback properties of OPA and the possibility of BWMOPO in NIMs were predicted in \cite{APB,OL,OL2}. Herein, we explore the effects of constructive and destructive nonlinear quantum interference processes in NLO centers embedded in NIMs on \emph{fully and quasi-resonant} four-wave mixing of the \emph{backward} waves that were not investigated in \cite{OL2}. Resonant coupling strongly enhances the NLO response of the composite. It appears that such a nanostructured composite exhibits extraordinary multiple-resonance properties with respect to the thickness of the slab, the density of the embedded centers, and the frequencies and strengths of the control fields. It also appears, that the linear phase-mismatch introduced by the host material can be negated at near-resonant FWM coupling of backward and ordinary waves. The investigation of such properties is important for the optimization of direct, coherent energy transfer from the control fields to the counter-propagating negative-index signal and the positive-index idler. This opens opportunities for the compensation of optical losses in NIMs and the creation of unique NIM-based photonic microdevices.

\section{Basic idea}
\begin{figure}[t]
\begin{center}
\includegraphics[width=.7\columnwidth]{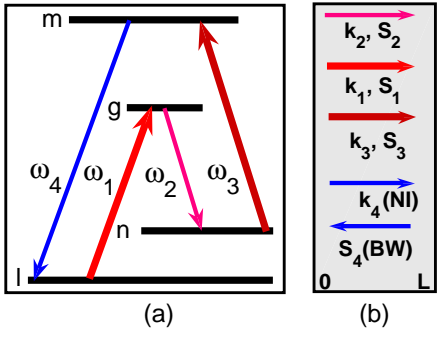}
\end{center}
\caption{Scheme of quantum controlled FWM interaction (a) and coupling geometry (b). Here, $\protect\omega_4$ is frequency of the signal, $\protect\omega_1$ and $\protect\omega_3$ of the control fields, $\protect\omega_2$ of the idler, $n(\protect\omega_4)< 0$.}
\label{f1}
\end{figure}

The basic scheme of resonant FWM of the BW in a NIM is as follows. A slab of NIM is doped by four-level nonlinear centers [Fig. \ref{f1}(a)] so that the signal frequency, $\omega_4$, falls in the NI domain [$n(\omega_4)< 0$], whereas all the other frequencies, $\omega_1$, $\omega_3$ and $\omega_2$, are in the the positive index (PI) domain. Below, we show the possibility to produce transparency and even amplification above the oscillation threshold at $\omega_4$ controlled by two lasers at $\omega_1$ and $\omega_3$ [Fig.~ \ref{f1}(b)]. These fields generate an idler at $\omega_2=\omega_3+\omega_1- \omega_4$, which experiences population inversion or Raman amplification caused by the control fields. This opens an additional channel of \emph{ energy transfer from the control fields to the signal}. The amplified idler contributes back to $\omega_4=\omega_3+\omega_1-\omega_2$ through FWM and thus causes strongly enhanced OPA of the signal. Unlike ordinary off-resonant NLO, a many-order resonance enhancement of the NLO coupling is followed by a laser-induced \emph{strong change of the local optical parameters}. The control fields may cause population transfer and even inversion, as well as modulation of the probability amplitude, and split and quantum interference modifications of the resonance shapes. Through these effects, therefore, the control fields may serve as a tool for harnessing the local optical coefficients. Alternatively, such changes can be minimized so that the major amplification would come directly from the energy exchange between the control fields, the signal and the idler through the FWM processes. In fact, the interference of quantum pathways in the vicinity of the resonances may even lead to possibility that the overall process ceases to be seen as a set of successive one- and multi-photon elementary processes
\cite{PRA} (and references therein).

First we shall show that the features of OPA in NIMs, which are employed in this work, appear in stark contrast with the basic properties of such processes generally known for ordinary PIMs. We assume that only $ \omega_4$ is close to the NI resonance. All other fields are far enough away to remain in the PI frequency domain. All wave vectors are presumed to be co-directed in order to ensure maximum phase matching. Because the energy flow $\mathbf{S_{4}}=(c/4\pi)[\mathbf{E}\times \mathbf{H}]$ does not depend on the sign of the refractive index, it appears contradirected to $\mathbf{k}_4$ and, therefore to all other wave vectors [Fig. \ref{f1} (b)]. This imposes extraordinary features on the nonlinear propagation of the coupled waves. The electric components of the waves and the nonlinear polarization are taken to be of the form
\begin{eqnarray}
{E}_{j}(z,t)&=&({E}_{j}/2)\exp [i(k_{j}z-\omega _{j}t)]+c.c.,  \label{ez} \\
P_{4,2}^{NL}(z,t)&=&(P_{4,2}^{NL}/2)\exp \{i[\widetilde{k}
_{4,2}z-\omega_{4,2}t]\}+c.c. \qquad  \label{p}
\end{eqnarray}
Here, $j=\{1-4\}$, $\omega_{4}+\omega_{2}=\omega_{3}+\omega_{1}$, $\widetilde{k}_{4,2}=k_{1}+k_{3}-k_{2,4}$, $k_{j}=|n_{j}|\omega_{j}/c>0$, $
P_{4,2}^{NL}/2=\chi_{4,2}^{(3)}E_{1}E_{3}E_{2,4}^{\ast }$ and $
\chi_{4,2}^{(3)}$ are effective nonlinear susceptibilities. Then the
slowly-varying amplitudes of the coupled waves at $\omega_4$ and $\omega_2$ are given by the equations
\begin{eqnarray}
{dE_{4}}/{dz} &=&-i\gamma_4^{(3)} E_{2}^{\ast }\exp [i\Delta kz]+({\alpha_{4}%
}/{2} )E_{4},  \label{4} \\
{dE_{2}}/{dz} &=&i\gamma_2^{(3)} E_{4}^{\ast }\exp [i\Delta kz]-({\alpha_{2}}%
/{2 })E_{2}.  \label{2}
\end{eqnarray}
Here, $\gamma_{2,4}^{(3)}= ({4\pi}|\mu_{2,4}|\omega^2_{2,4}/k_{2,4}c^2) \chi_{2,4}^{(3)}E_{1}E_{3}$ are NLO coupling coefficients; $\epsilon_j$ and $ \mu_j$ are the dielectric permittivities and magnetic permeabilities (which are negative at $\omega_4$); $\Delta k=k_{1}+k_{3}-k_{2}-k_{4}$; and $ \alpha_{j}$ are the absorption or amplification coefficients. The amplitudes of the fundamental (control) waves $E_{1}$ and $E_{3}$ are assumed constant along the slab. Note that there are \emph{three fundamental differences} in equations (\ref{4} ) and (\ref{2}) as compared with their counterparts in ordinary materials with positive permeability $\mu$ and permittivity $\epsilon$ at all frequencies. First, the sign of the nonlinear polarization term $ \gamma_{4}^{(3)}$ is opposite to that of $\gamma_{2}^{(3)}$, which occurs because $\mu_{4}<0$. Second, the opposite sign appears for $\alpha_{4}$. Third, the boundary conditions for $a_{4}$ must be defined at the opposite side of the sample as compared to those for all other waves. These differences lead to a counterintuitive evolution of the signal and idler along the medium. Assuming $E_{1,3}$ constant and taking into account the boundary conditions $E_{4}(L)=E_{4L}$ and $E_{2}(0)$ = 0 (where $L$ is the slab thickness), the solutions to equations (\ref{4})-(\ref{2}) are found as
\begin{eqnarray}
E_{4}(z) &=&A_{1}\exp [(\beta_{1}^+z]+ A_{2}\exp [(\beta_{2}^+z],
\label{e4z} \\
E_{2}^{\ast }(z) &=&\kappa_{1}A_{1}\exp [\beta_{1}^-z]+ \kappa_{2}A_{2}\exp
[\beta_{2}^-z],  \label{e2z}
\end{eqnarray}
where,
\begin{eqnarray}  \label{not}
&\beta_{1,2}^{\pm}=\beta_{1,2}\pm (i\Delta k/2), \,
\beta_{1,2}=(\alpha_{4}-\alpha_{2})/(4)\pm iR, \,  \nonumber \\
&A_{1}={E_{4L}\kappa_{2}}/D, \, A_{2}=-{E_{4L}\kappa_{1}}/D,  \nonumber \\
&D=\kappa_{2}\exp [\beta_{1}^+L]-\kappa_{1}\exp [\beta_{2}^+L],  \nonumber \\
&\kappa_{1,2}=(\pm {R}+is)/\gamma_4,\, R=\sqrt{g^2-s^{2}},  \nonumber \\
&g^2=\gamma_2^{\ast}\gamma_4,\, s=({\alpha_{4}+\alpha_{2}})/{4}-i({\Delta k}/%
{2}).  \nonumber
\end{eqnarray}
The transmission (amplification) factor for the negative-index signal, $T_4(z)=\left\vert {E_{4}(z)}/{E_{4L}}\right\vert ^{2}$, is given by the equation
\begin{equation}
T_4(0)=T_{40}=\left|\frac{\exp \left\{-\left[ \left( \alpha_{4}/2\right)-s%
\right] L\right\}}{\cos RL+\left(s/R\right) \sin RL}\right|^2.  \label{T}
\end{equation}
The fundamental difference between the spatial dependence of OPA in ordinary and NIM materials is explicitly seen at $\alpha_j=\Delta k=0$ in the off-resonant case, where $\gamma_2^{\ast}=\gamma_4=g$ and hence $g^2$ is a real, positive quantity. Then equations~(\ref{e4z})-(\ref{not}) reduce to
\begin{equation}
T_4(z)=\left|{\cos(g z)}/{\cos(g L)}\right|^2,  \label{T1}
\end{equation}
whereas in ordinary media the signal would exponentially grow as $T\propto \exp(2gz)$ \cite{Yar}. Equation~(\ref{T1}) and output magnitude $T_{40}=\left|\cos(g L)\right|^{-2}$ present a sequence of \emph{geometrical} resonances, with behavior similar to that of distributed-feedback resonance. Thus, in contrast with ordinary media, the spatial distribution of the signal inside the slab and its output value at $z=0$ exhibit unusual oscillatory dependence on the slab thickness. Such extraordinary resonances provide for the feasibility of attaining the \emph{oscillation threshold} for the generation of \emph{the entangled counter-propagating left-handed,
$\hbar\omega_4$, and right-handed, $\hbar\omega_2$, photons without a cavity}. Such generation would occur when $g L \rightarrow (2j+1)\pi/2$. The rate of the energy transfer from the control fields to the signal depends on the local strengths of both the signal and the idler. Thus, a change in their absorption and amplification rates my result in a dramatic change in distribution of the signal and the idler across the slab and, consequently, in their output values.  Ultimately, the tailoring of the output signal stems from its extraordinary evolution along the slab.
\begin{figure}[h!]
\begin{center}
\includegraphics[height=.55\columnwidth]{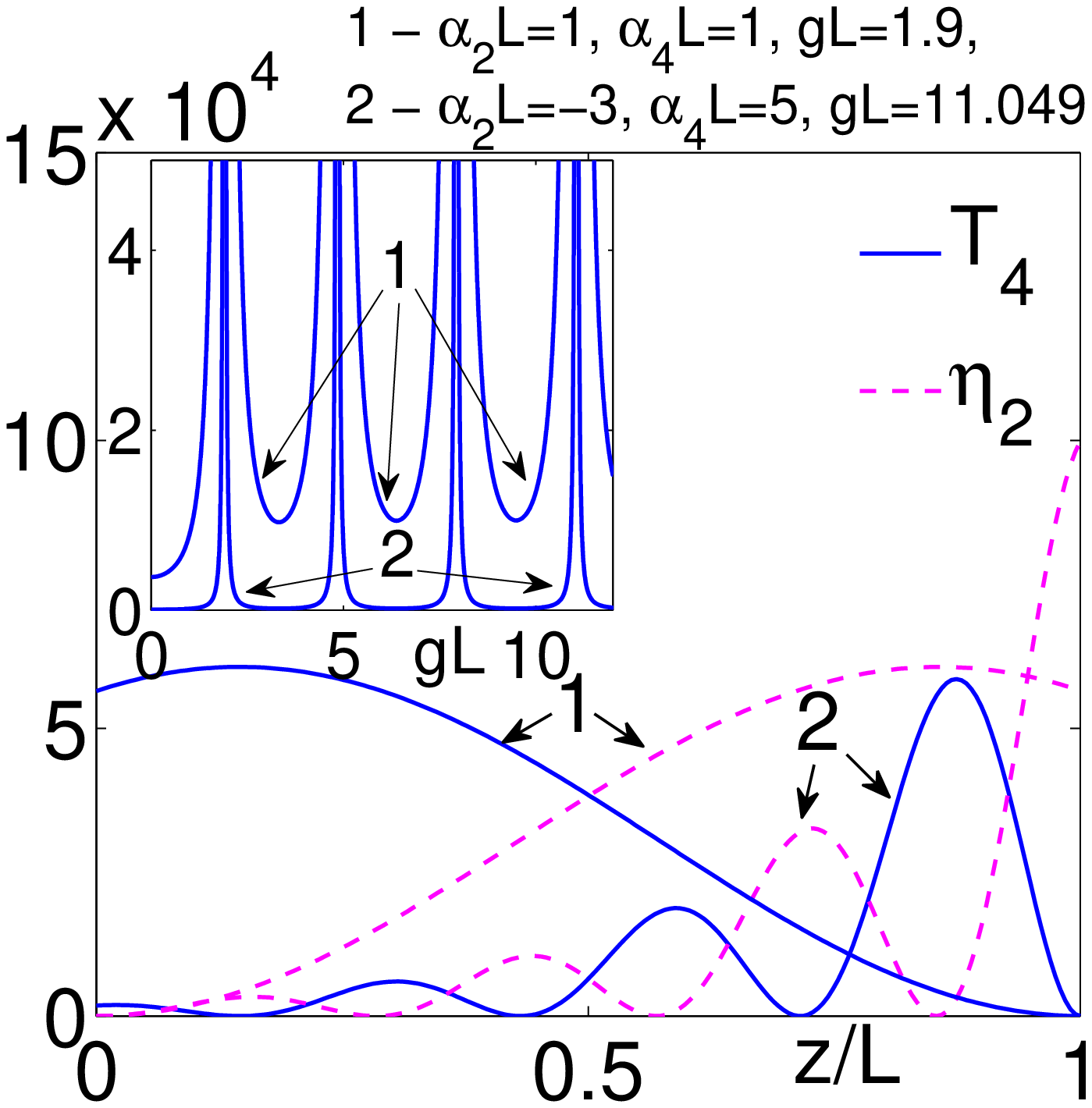}
\includegraphics[height=.55\columnwidth]{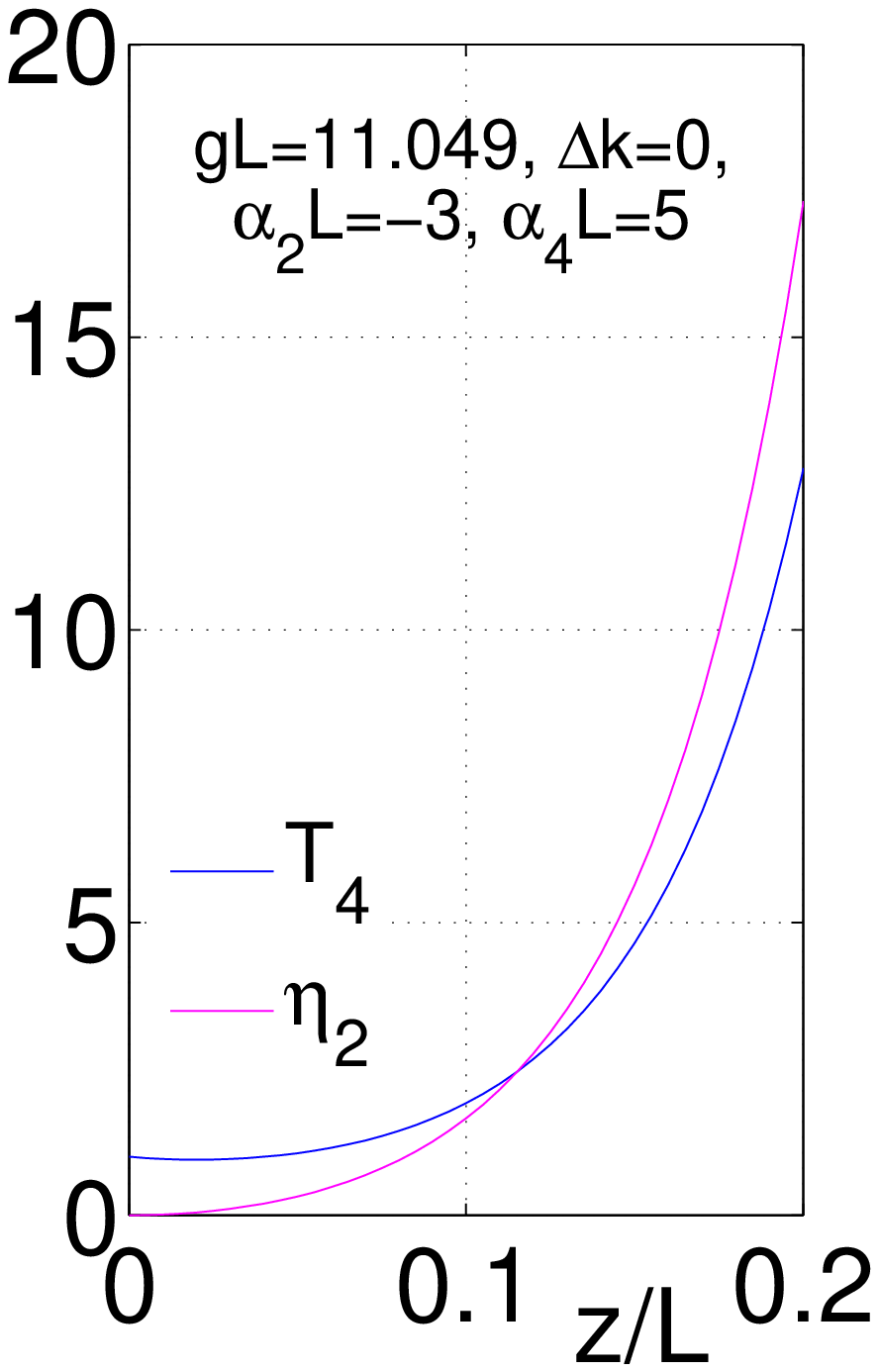}
\\[0pt]
\hspace{15mm}(a) \hspace{30mm} (b)
\end{center}
\caption{\label{f2} The effect of amplification of the idler and the difference between the distribution of the signal, $T_{4}(z)$,
and the idler, $\protect\eta_2=|E_{2}(z)/E_{4}(L)|^2$, along the slab NIM, main plots (a); and in ordinary materials, (b),  at otherwise identical conditions. Inset (a): geometrical resonances in output signal $T_4(z=0)$. 1: $\alpha_2L=\alpha_4L=1$, gL=1.9; 2: $\alpha_2L=-3$, $\alpha_4L=5$, gL=11.049.}
\end{figure}

An example of such dependence and its difference from that in the ordinary materials is depicted in Fig.~\ref{f2}(a),(b). The  inset in Fig.~\ref{f2}~(a) shows narrow resonances in the output signal at $z=0$ as the function of the quantity gL. The position and shape of the resonances depend on the absorption or amplification indices. The main plot shows the oscillation behavior of the signal and the idler across the slab at a slightly off-resonant values of $gL$, which results in the overall depletion of the signal and the idler. Since the idler grows toward the back facet of the slab at $z=L$ and the signal experiences absorption or amplification in the opposite direction, the maximum of the signal at given parameters may appear closer to the back facet of the slab. As seen from the comparison of plots 1 and 2 in Fig.~\ref{f2}~(a), such distribution depends strongly on the difference in the absorption or amplification rates for the signal and the idler.
Hence, the changes in the slab thickness, or in the intensity of the control fields, or in their absorption indices lead to significant, resonance,  changes  in the output signal. Such a dependence is in stark contrast with its counterpart in positive-index materials shown in Fig.~\ref{f2}~(b).

\section{Resonant coupling and quantum control: numerical simulations}
This main section is aimed at numerical experiments to show specific the counterintuitive features and data which are characteristic for resonant and near-resonant four-wave coupling of backward waves. The information presented here can be used for the all-optical tailoring of such coupling.

\subsection{Quasi-resonant coupling and compensation of phase mismatch} An essential difference of the \emph{resonant} nonlinear processes under investigation from ordinary off-resonant coupling is that the nonlinear susceptibilities $\gamma_4$ and $\gamma_2$ become complex and different from each other in this case.
\begin{figure}[!h]
\begin{center}
\includegraphics[width=.48\columnwidth]{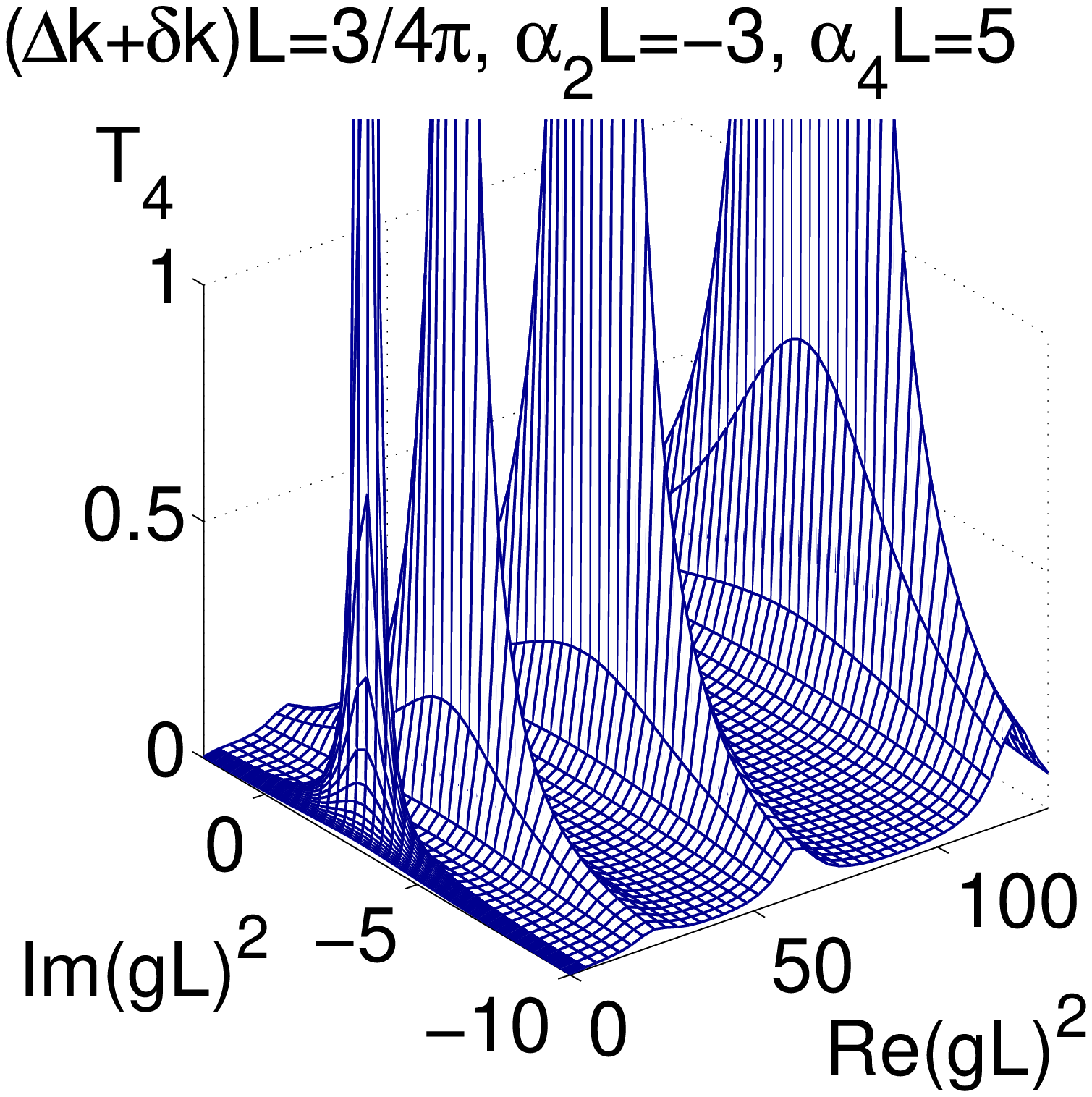}
\includegraphics[width=.48\columnwidth]{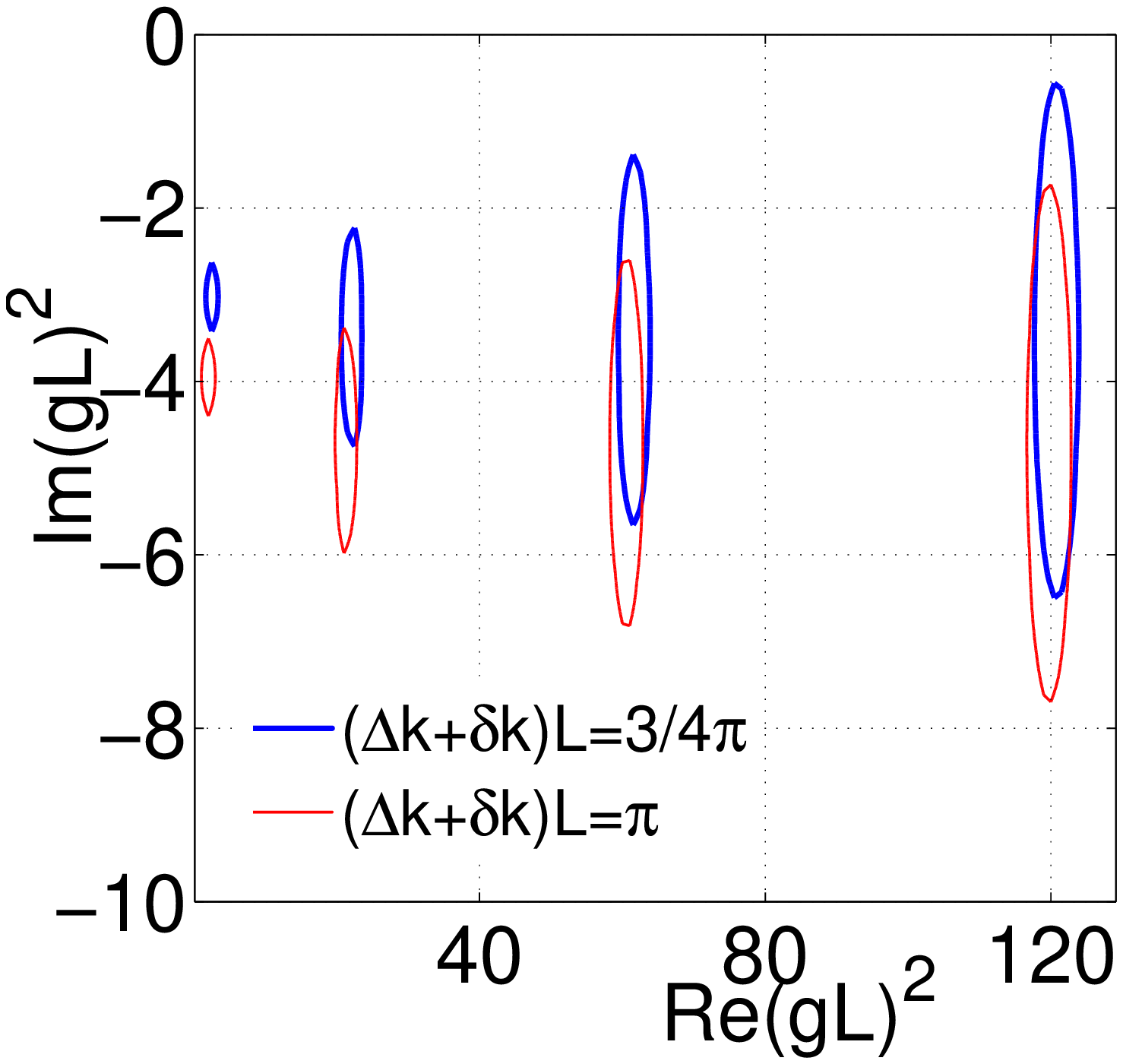}
\\[0pt]
(a) \hspace{30mm} (b)\\[0pt]
\end{center}
\caption{\label{f3} Correlation between the transmittance $T_4$ of the NIM slab, $\Re(gL)$, $\Im(gL)$ and combined linear phase mismatch introduced by the host material, $\protect\delta k$, and doping material, $\Delta k$. $\protect\alpha_2L=-3$, $\protect\alpha_4L=5$. a: $(\Delta k+\protect\delta k)L=(3/4)\pi$. (b): $T_4=1$; solid line -- phase mismatch $(\Delta k+\protect\delta k)L=(3/4)\pi$, dashed line -- phase mismatch $(\Delta k+\protect\delta k)L=\pi$.
}
\end{figure}
Hence, the factor $g^2$ may become negative or complex, which causes additional \emph{radical changes} in the nonlinear propagation features. Figure~\ref{f3} shows that resonant coupling and, hence, complex or negative magnitudes of $g^2$ may lead to the result that the requirement $\Delta k+\delta k=0$ \emph{ceases to be optimum} for the process under investigation, which is also in striking contrast with the off-resonant OPA of the weak signal in ordinary materials. Some mismatch $\delta k$, introduced by the host material, may bring the optimum signal frequency to the maximum of the NLO conversion rate. Alternatively, transparency and oscillation may become impossible for the frequencies that correspond to maximum nonlinear response unless the phase mismatch is elaborately added. Specifically, Fig.~\ref{f3}(a) and Fig.~\ref{f3}(b) prove the possibility of negating of the combine phase mismatch $(\Delta k+\protect\delta k)L=(3/4)\pi$ and $(\Delta k+\protect\delta k)L=\pi$. Larger absolute values of negative $\Im(gL)^2$ provide for compensation of larger positive values of $(\Delta k+\protect\delta k)L$. Note, however, that the requirement of linear phase mismatch is known for coupling strong fields due to the fact that the ratio of the intensities of the coupled fields and, consequently, the coupling phase varies along the medium. A similar situation occurs here for the weak counter-propagating waves and the control fields that are homogeneous along the slab.

\subsection{Numerical model for embedded resonant NLO centers}
To demonstrate the outlined extraordinary NLO features and to prove the proposed possibilities of tailored transparency and oscillations, we have adopted the following model for numerical simulations, which is characteristic for molecules imbedded in a solid host: the energy level relaxation rates are $\Gamma_n=20 \times 10^6$~s$^{-1}$, $ \Gamma_g=\Gamma_m=120 \times 10^6$~s$^{-1}$; the partial transition probabilities are $\gamma_{gl}=7\times 10^6$~s$^{-1}$, $\gamma_{gn}=4\times 10^6$~s$^{-1}$, $\gamma_{mn}=5\times 10^6$~s$^{-1}$, $\gamma_{ml}=10\times 10^6$~s$^{-1}$; and the homogeneous transition half-widths are $ \Gamma_{lg}=10^{12}$~s$^{-1}$, $\Gamma_{lm}=1.9 \times 10^{12}$~s$^{-1}$, $\Gamma_{ng}=1.5 \times 10^{12}$~s$^{-1}$, $\Gamma_{nm}=1.8 \times 10^{12}$~s$ ^{-1}$, $\Gamma_{gm}=5 \times 10^{10}$~s$^{-1}$; $\Gamma_{ln}=10^{10}$~s$ ^{-1}$. We assume that $\lambda_2=756$ nm and $\lambda_4=480$ nm. The density-matrix method described in \cite{PRA} is used for calculation of the intensity-dependent local parameters, and we also included quantum nonlinear interference effects. This allows us to account for changes in absorption, amplification and refractive indices as well as accounting for the striking changes in the magnitudes and signs of the NLO susceptibilities. The parameters $\Re\gamma_{4,2}$, $\Im\gamma_{4,2}$ and, hence, $\Re g^2$ and $ \Im g^2$ may experience significant changes also due to the population redistribution over the coupled levels caused by the control fields. The latter strongly depends on the ratio of the partial transition probabilities. In the numerical experiments below, the coupling Rabi frequencies that represent the strength of the control fields are introduced as $G_{1}= E_1d_{lg}/2\hbar$, $G_{3} = E_3d_{nm}/2\hbar$. For the optical transitions indicated above, the magnitudes $G \sim 10^{12}$~s$^{-1}$ correspond to the control field intensities in the range of $I \sim$ 10 - 100 kW/(0.1mm)$^2$. The resonance offset is $\Omega_4=\omega_4-\omega_{ml}$; other resonance detunings $\Omega_{j}$ are defined in a similar way. The values $\chi_{4,2}$ denote the effective linear susceptibilities, while $ \alpha_{40}$ and $\chi_{40}$ are their fully resonant values with all driving fields turned off. We assumed that 90\% of the signal is absorbed by the host slab.

\subsection{Fully resonant control fields}
\begin{figure}[h]
\begin{center}
\includegraphics[width=.325\columnwidth]{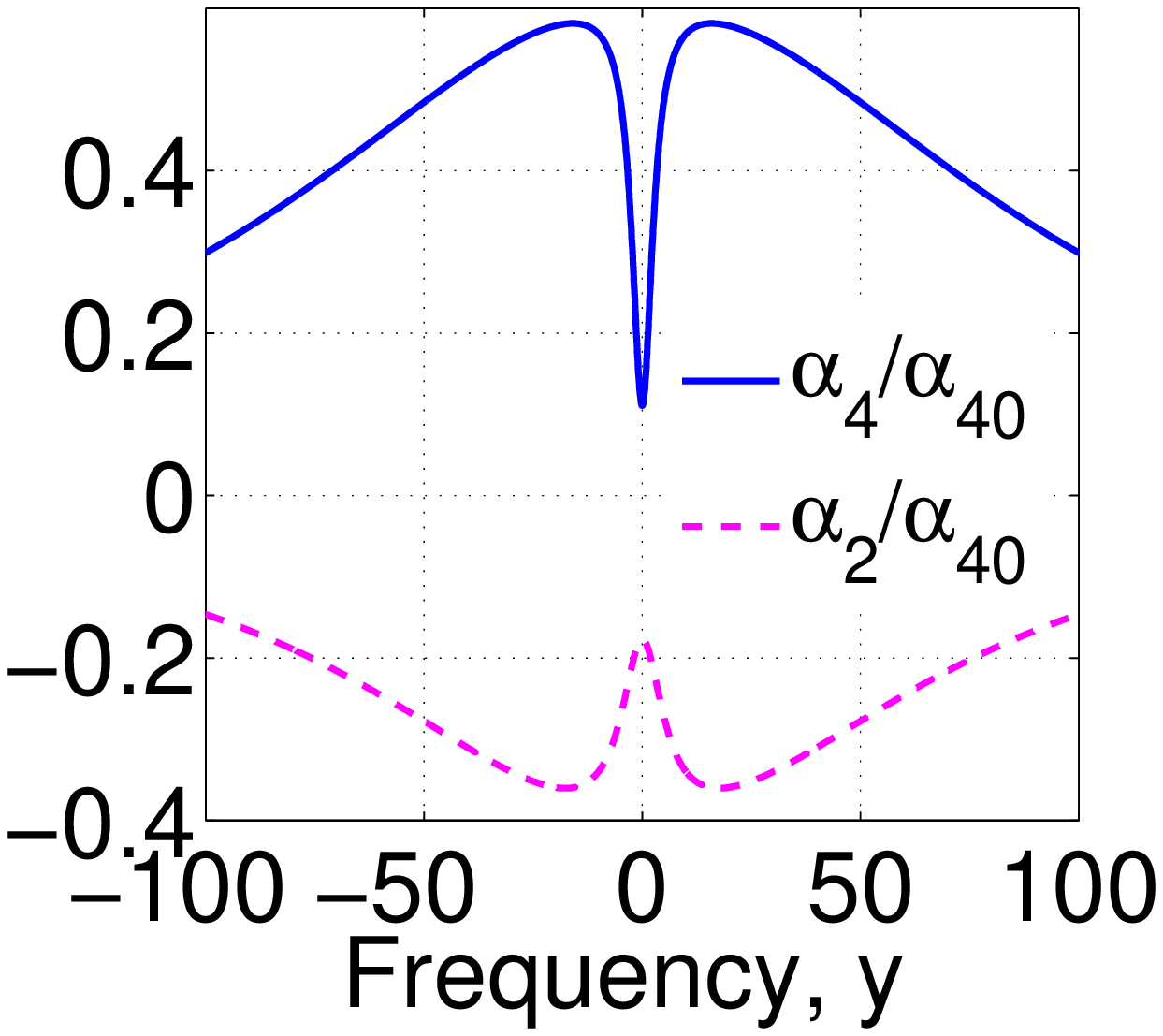}
\includegraphics[width=.325\columnwidth]{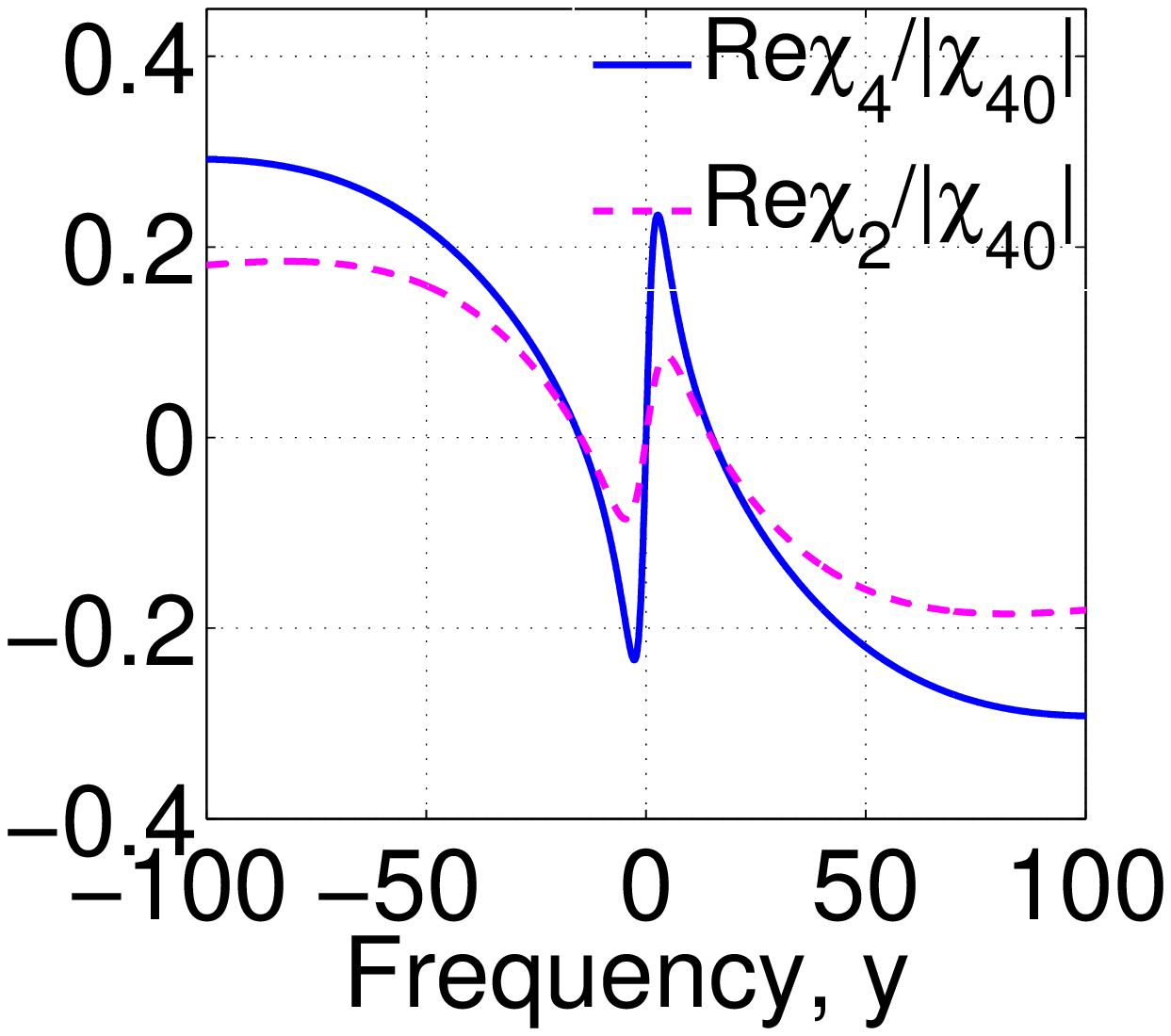}
\includegraphics[width=.325\columnwidth]{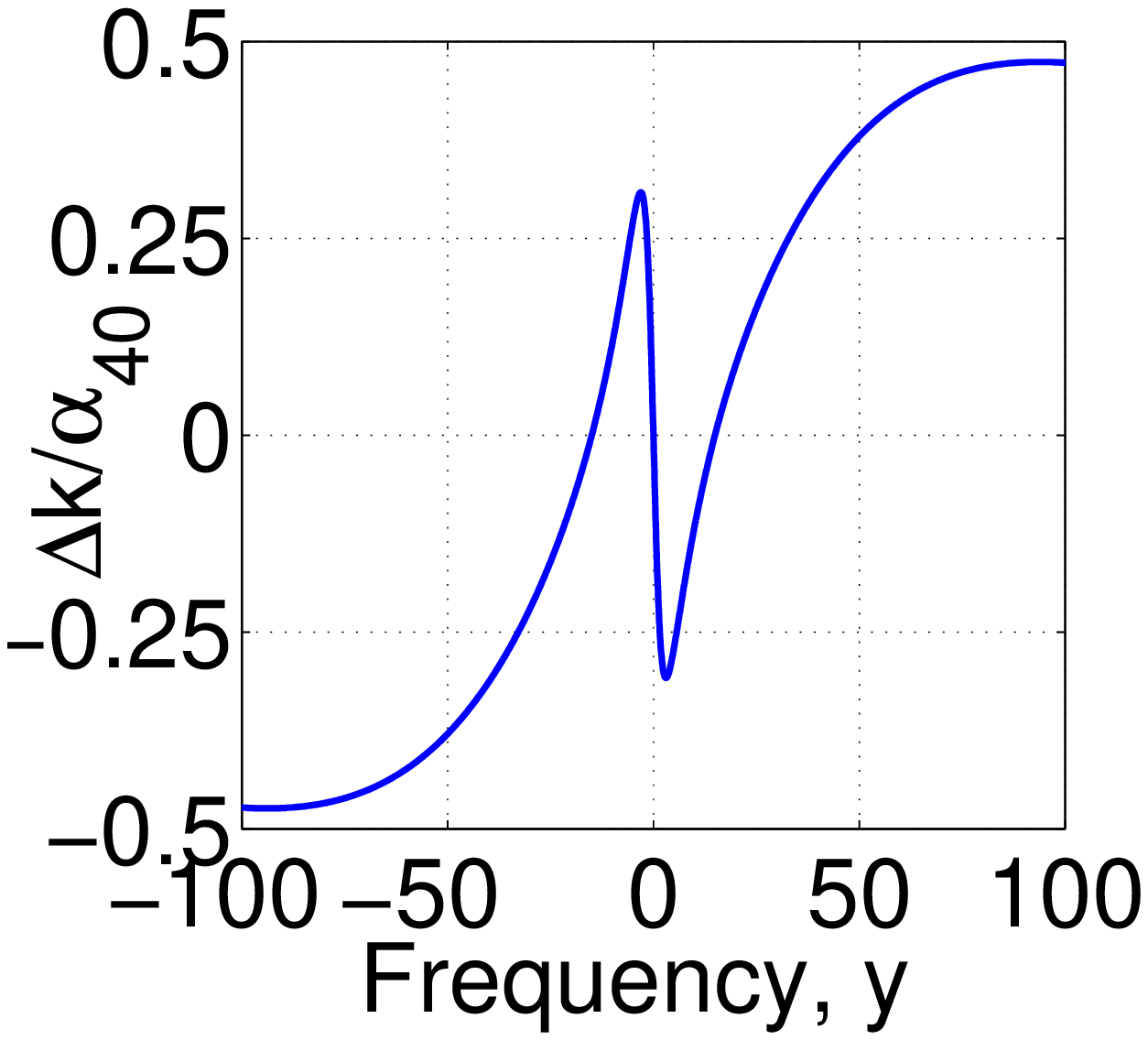}
\\[0pt]
(a) \hspace{20mm} (b)\hspace{20mm} (c)\\[0pt]
\includegraphics[width=.325\columnwidth]{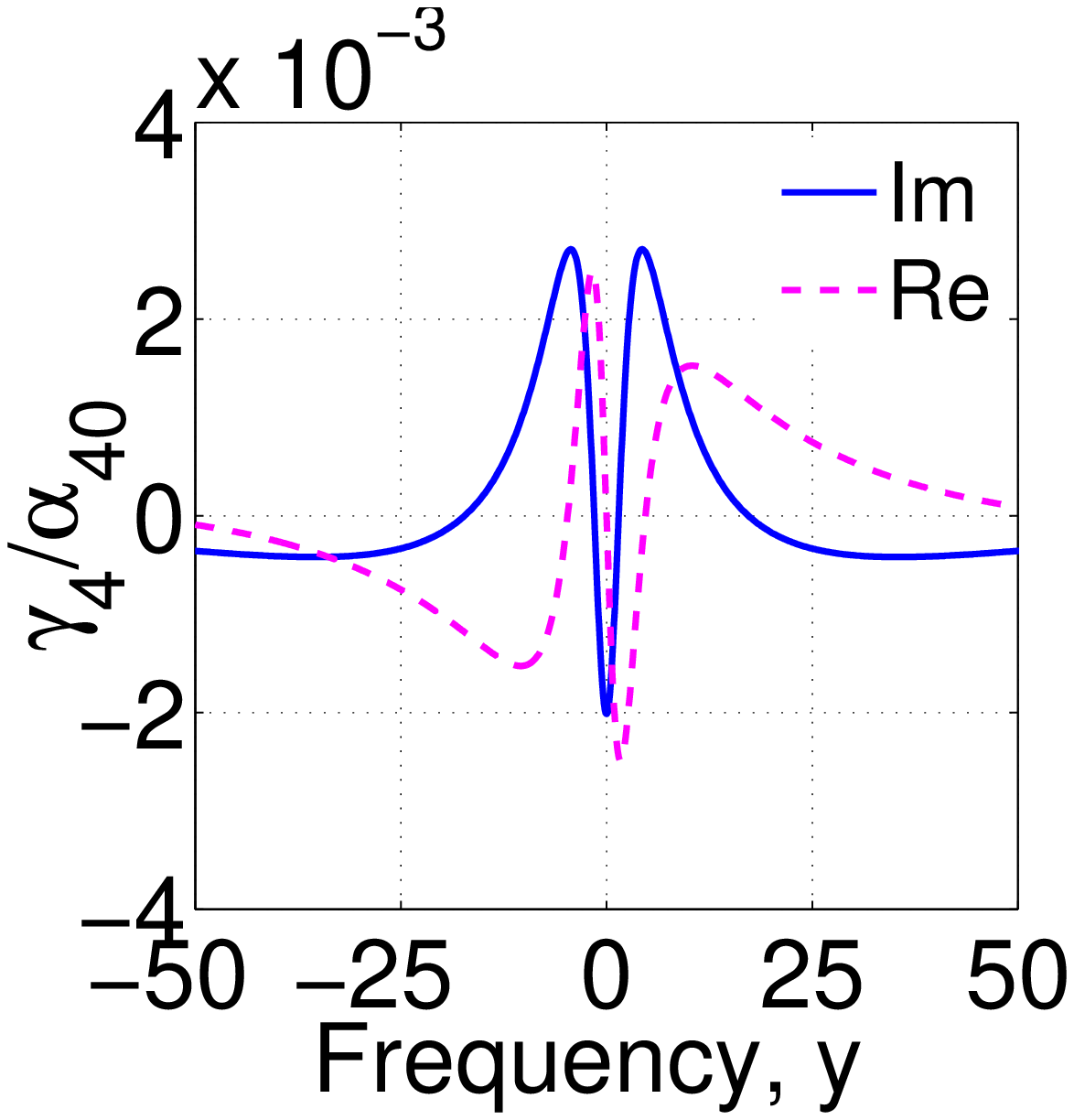}
\includegraphics[width=.325\columnwidth]{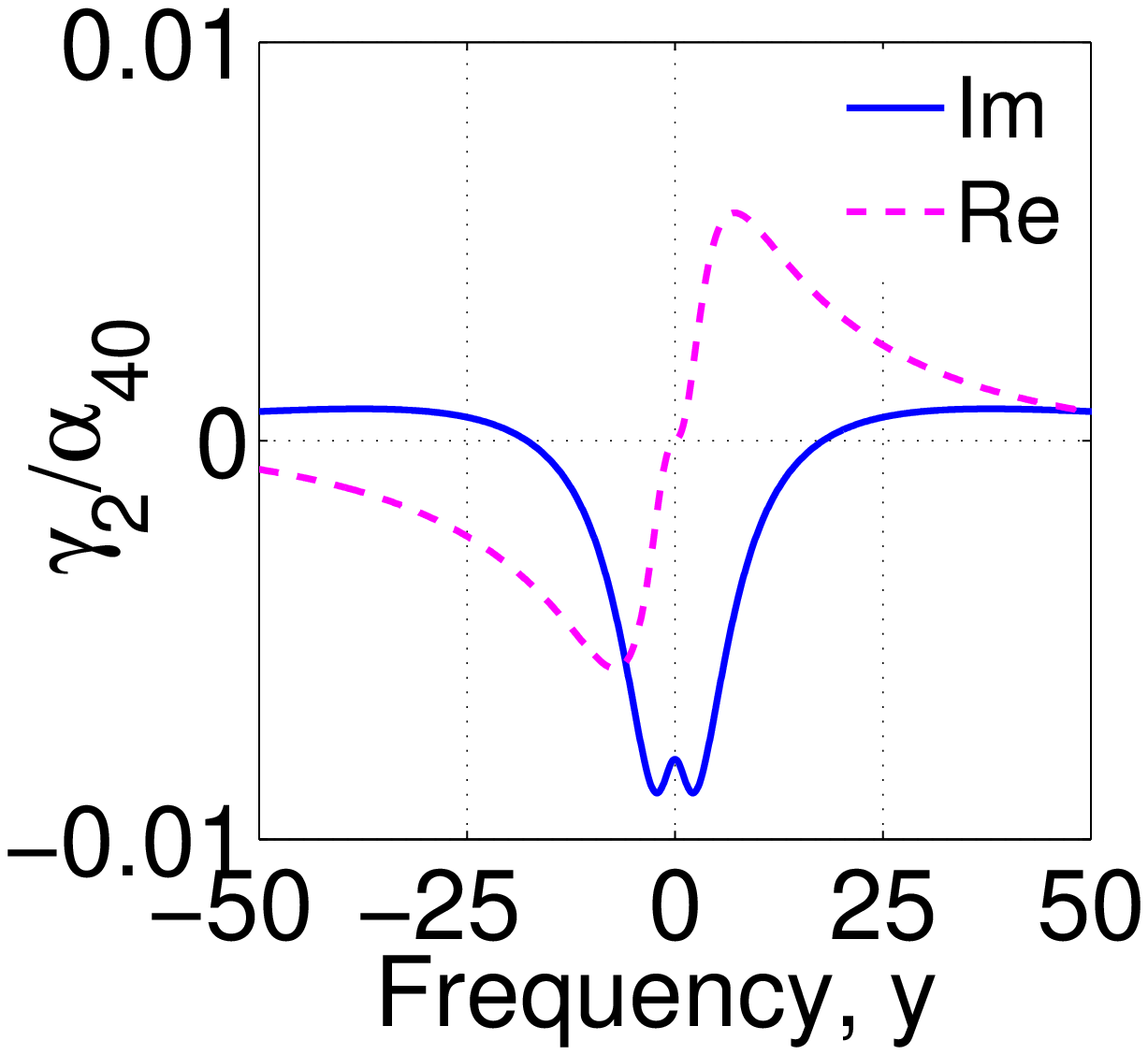}
\includegraphics[width=.325\columnwidth]{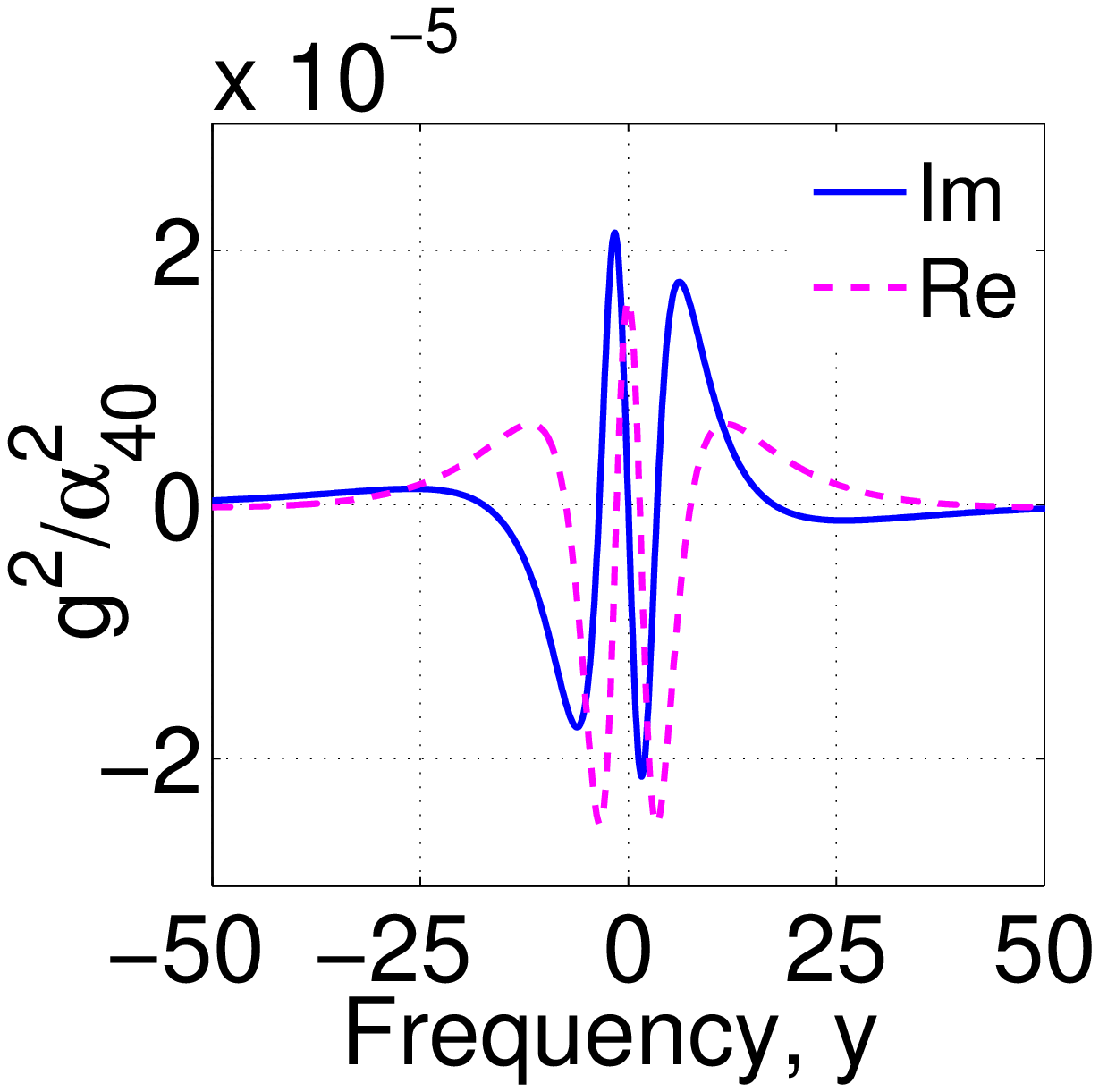}
\\[0pt]
(d) \hspace{20mm} (e)\hspace{20mm} (f)\\[0pt]
\end{center}
\caption{\label{f4} Nonlinear interference structures in the local optical parameters attributed to the embedded centers. $y=\Omega_4/(0.01 \Gamma_{lm})$. $\Omega_1=\Omega_3=0$, $G_1=1.646$~GHz, $G_3=45.539$~GHz.
 }
\end{figure}

Figure~\ref{f4} depicts the local parameters entering in Eq.~(\ref{T}) with fully resonant control fields at the given strengths. All parameters are scaled to the indicated fully resonant values taken at all strong fields turned off. The quantity $\Re\chi_4$ represents changes in the refractive index at $\omega_4$. 
Numerical analysis is done with the aid of the steady-state solutions to the density matrix equations \cite{PRA}. Here, $y_s=\Omega_4/(0.01 \Gamma_{lm})$. Modulation of the quantum amplitudes by the strong driving fields and constructive and destructive interference of multistep and multiphoton quantum pathways at the coupled transitions cause the appearance of \emph{nonlinear interference spectral structures} in the effective local linear and nonlinear optical parameters \cite{PRA} (and references therein). Figure~\ref{f4} shows that the changes caused by the control fields in the  local optical parameters may be commensurable with their original values. Consequently, the nonlinear propagation of the signal and idler can be tailored through such processes. Control fields at given intensities cause appreciable redistribution of the energy-level populations accompanied by  population inversion at the resonant idler transition:
$r_l\approx 0.60$, $r_g\approx 0.38$, $r_n\approx 0.01$,
$r_m\approx 0.01$. Other optical parameters at $\omega_4=\omega_{ml}$ are as follow:
$\alpha_4/\alpha_{40}\approx 0.11$,
$\alpha_2/\alpha_{40}\approx-0.18$,
$\Delta k/\alpha_{40}\approx 0.28$,
$\Im(\gamma_2/\alpha_{40})\approx-0.008$,
$\Re(\gamma_2/\alpha_{40})=0$,
$\Im(\gamma_4/\alpha_{40})\approx-0.002$,
$\Re(\gamma_4/\alpha_{40})=0$,
$\Im(g^2/\alpha_{40}^2)=0$,
$\Re(g^2/\alpha_{40}^2)\approx 1.6\times10^{-5}$.

\begin{figure}[!h]
\begin{center}
\includegraphics[width=.4\columnwidth]{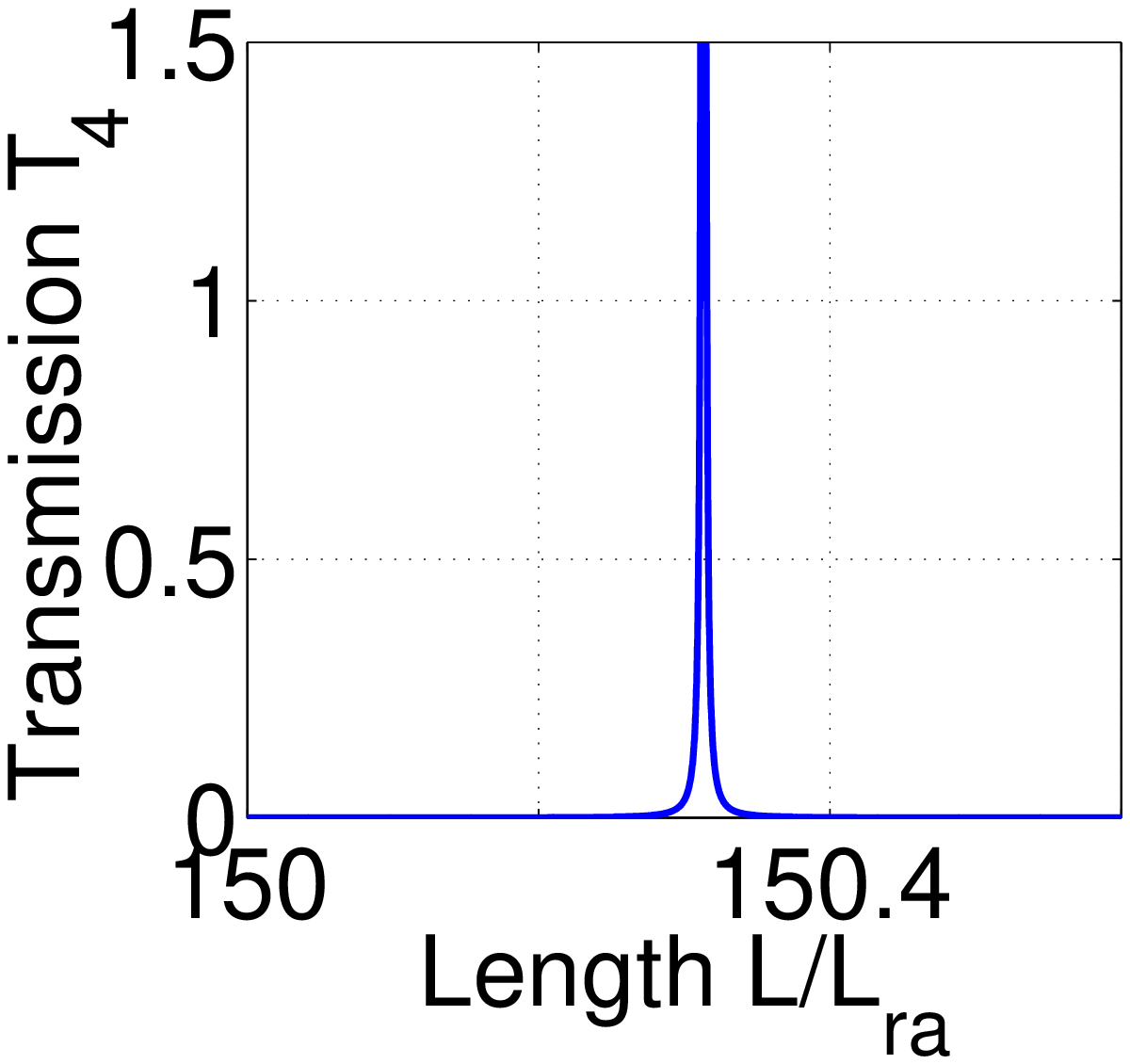}
\includegraphics[width=.4\columnwidth]{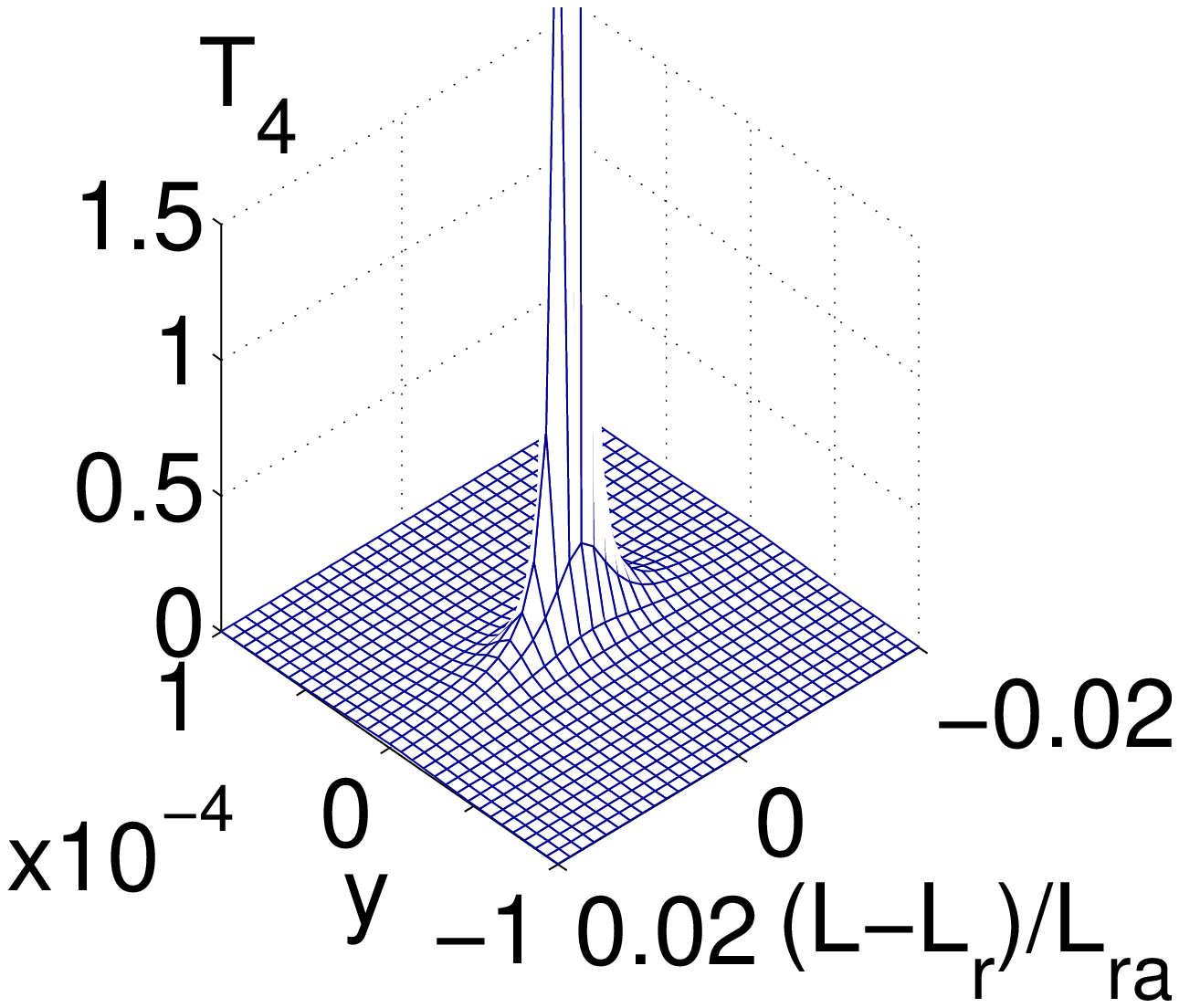}
\\[0pt]
(a) \hspace{30mm} (b)\\[0pt]
\includegraphics[width=.4\columnwidth]{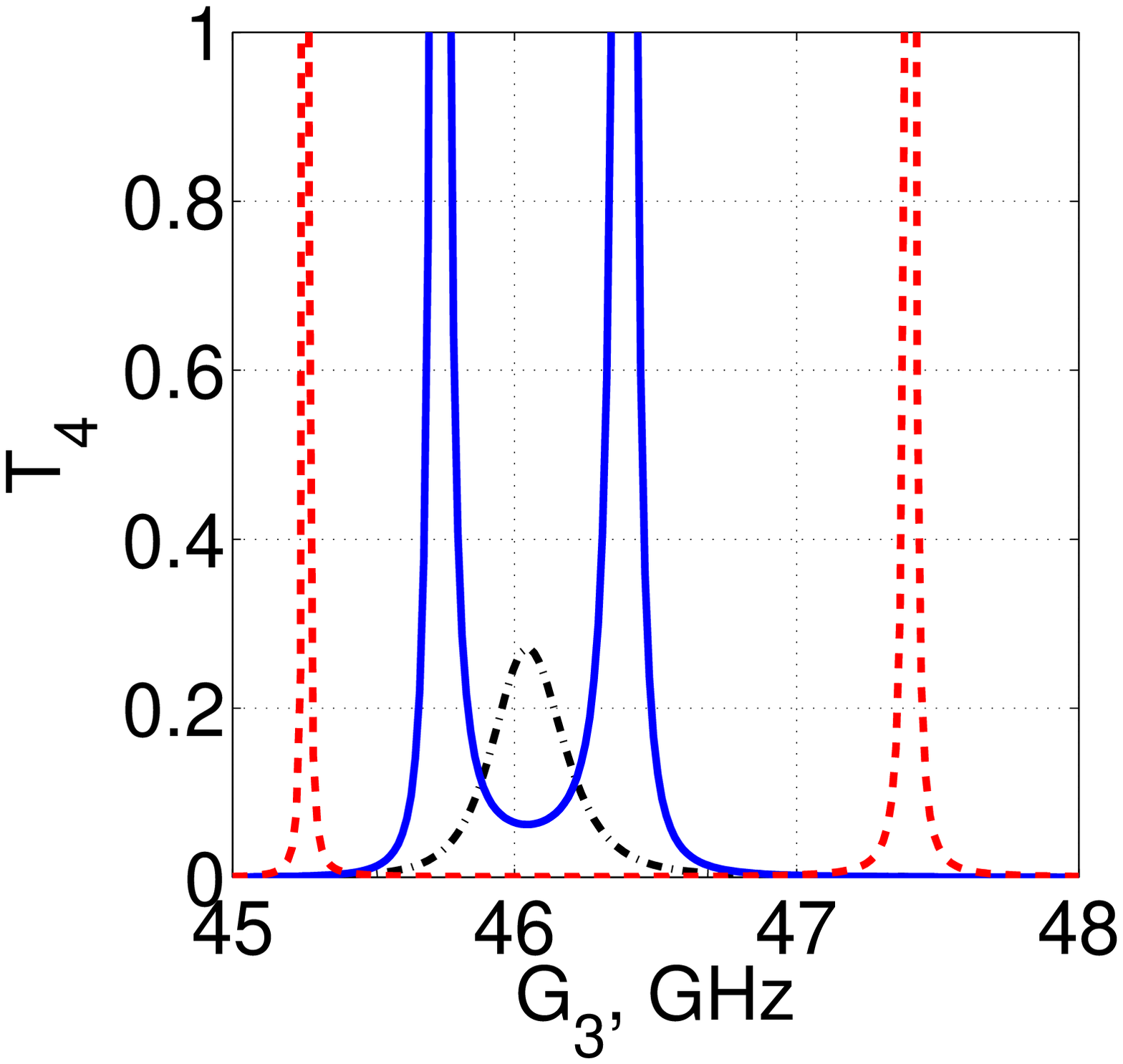}
\includegraphics[width=.4\columnwidth]{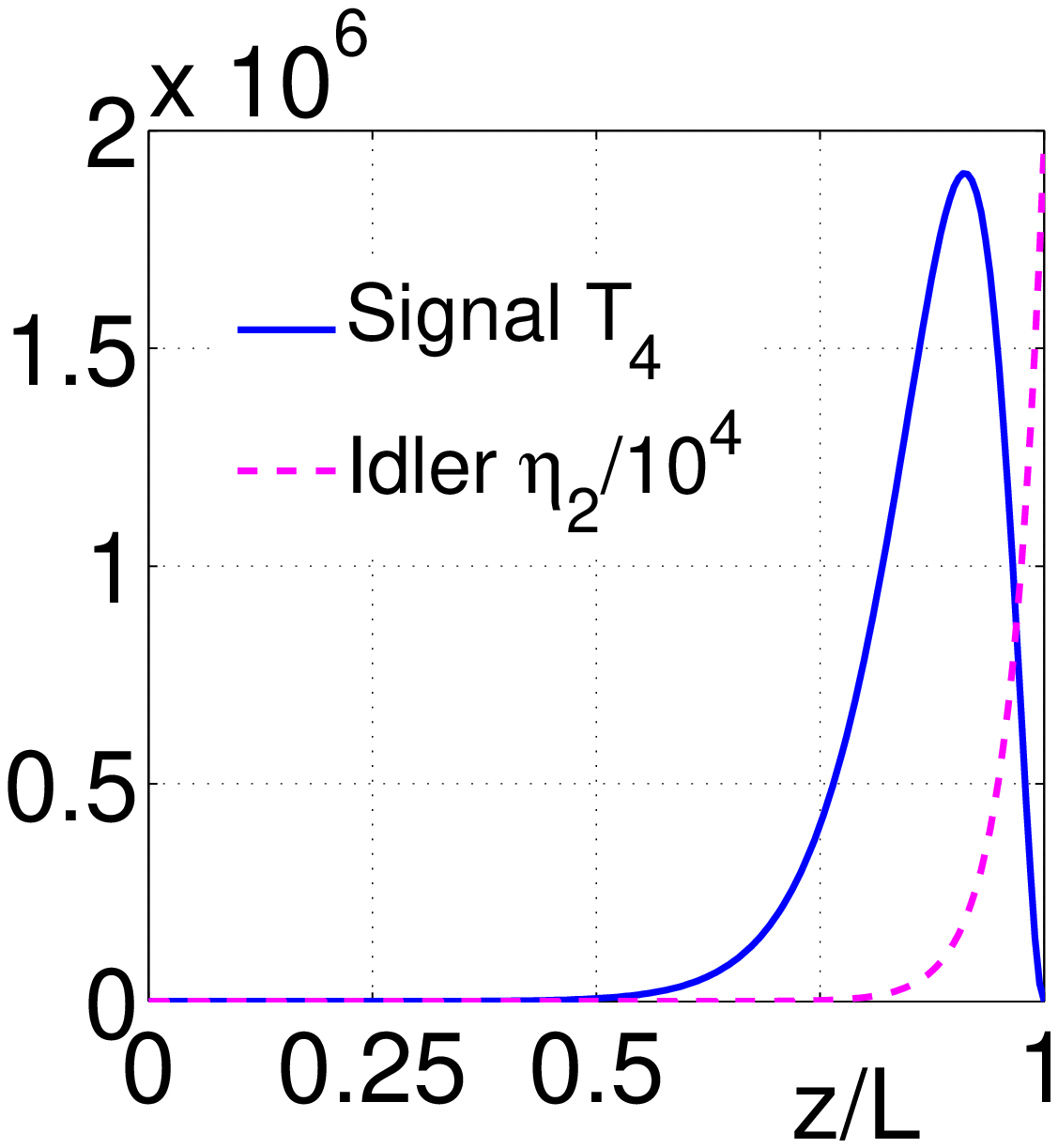}
\\[0pt]
(c) \hspace{30mm} (d)
\end{center}
\caption{\label{f5} Geometrical resonances in the output signal and idler, (a) and (b); dependence of the output signal on the intensity of the control fields at L=L$_r$, (c); and distribution of the signal and the idler inside the slab at L slightly different from L$_r$, (d).  $y=\Omega_4/(0.01 \Gamma_{lm})$.  $\Omega_1=\Omega_3=0$,  $\protect\alpha_{NIM}L=2.3$. (a), (c) and (d): $\protect\omega_4=\protect\omega_{lm}$. (a), (b) and (d): $G_1=1.646$~GHz, $G_3=45.539$~GHz.  (c): $L_r/L_{ra}=150.3130$, single peak (dash-dot)~-- $G_1=1.655$~GHz, solid line --~ $G_1=1.657$~GHz, dashed line~-- $G_1=1.678$~GHz; (d): $L/L_{ra}=150.31$.}
\end{figure}

Figure~\ref{f5} displays the resonance dependence of the output signal on the optical thickness of the slab, on the resonance offset of the signal, on the strengths of the control fields,  and
the distribution of the signal and idler along the slab as tailored by the given control fields.
As discussed above, the output signal for the waves coupled in the NIM slab through OPA exhibits a set of distributed feedback-type resonances. Such resonant behavior appears with respect to the intensity of the fundamental fields, as well as with respect to the product of the slab length and the density of the nonlinear centers (the optical density), and also with respect to the resonance offsets for the signal and fundamental fields. Optimization of the output signal at $z=0$ is determined by the interplay of absorption, idler gain, FWM and, hence, by the wave vector mismatch. This appears to be a multi-parameter problem associated with sharp resonance dependencies. Figure~\ref{f5} displays the results of the numerical analysis of Eq.(\ref{T}) based on the steady-state solutions to the density-matrix equations for one of the transmission resonances. In the given case, the strongest and most easily achieved resonance corresponds to the center of the signal transition at $\omega_4=\omega_{ml}$. Here, we introduce the scaled product of the slab length and the number density of the embedded centers, $L/L_{ra}$, through the resonance absorption length, $ L_{ra}=\alpha_{40}^{-1}$. Figure~\ref{f5}(a) displays a narrow geometrical resonance as a function of the slab thickness or the density of the embedded NLO centers at $L=L_r=150.3130$~$L_{ra}$ for the optimum frequency offset for the signal at $\Omega_4=0$. Figure~\ref{f5}(b) shows that the width of the transparency window is on a scale less than the narrowest (here Raman) transition half-width and the resonant absorption length. Figure~\ref{f5}(c) shows that the output signal at the given frequency and optical density of the slab is strongly dependent on the magnitude and ratio of the control field intensities. Figure~\ref{f5}(d) shows that the signal intensity inside the slab may significantly exceed its output value at $z$=0, and the idler intensity substantially exceeds that for the signal. Such features depend on the ratio of the OPA, absorption rate of the signal and amplification rate of the idler. Here, the slab thickness (or density of the nonlinear centers) is taken to be slightly different from its resonance value. Calculated

\subsection{Frequency offsets on the order of the narrowest Raman transition width}
\begin{figure}[h!]
\begin{center}
\includegraphics[width=.325\columnwidth]{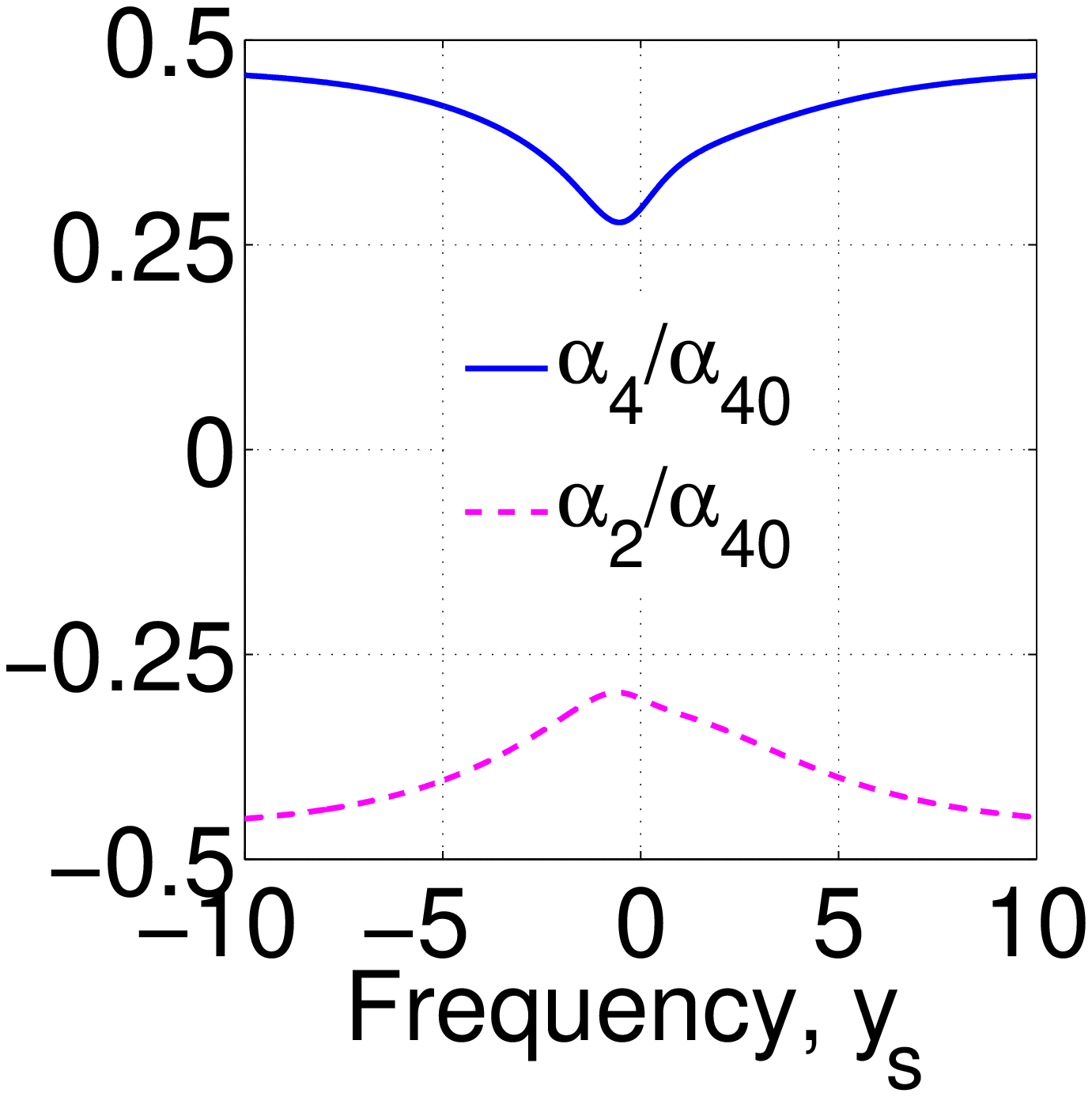}
\includegraphics[width=.325\columnwidth]{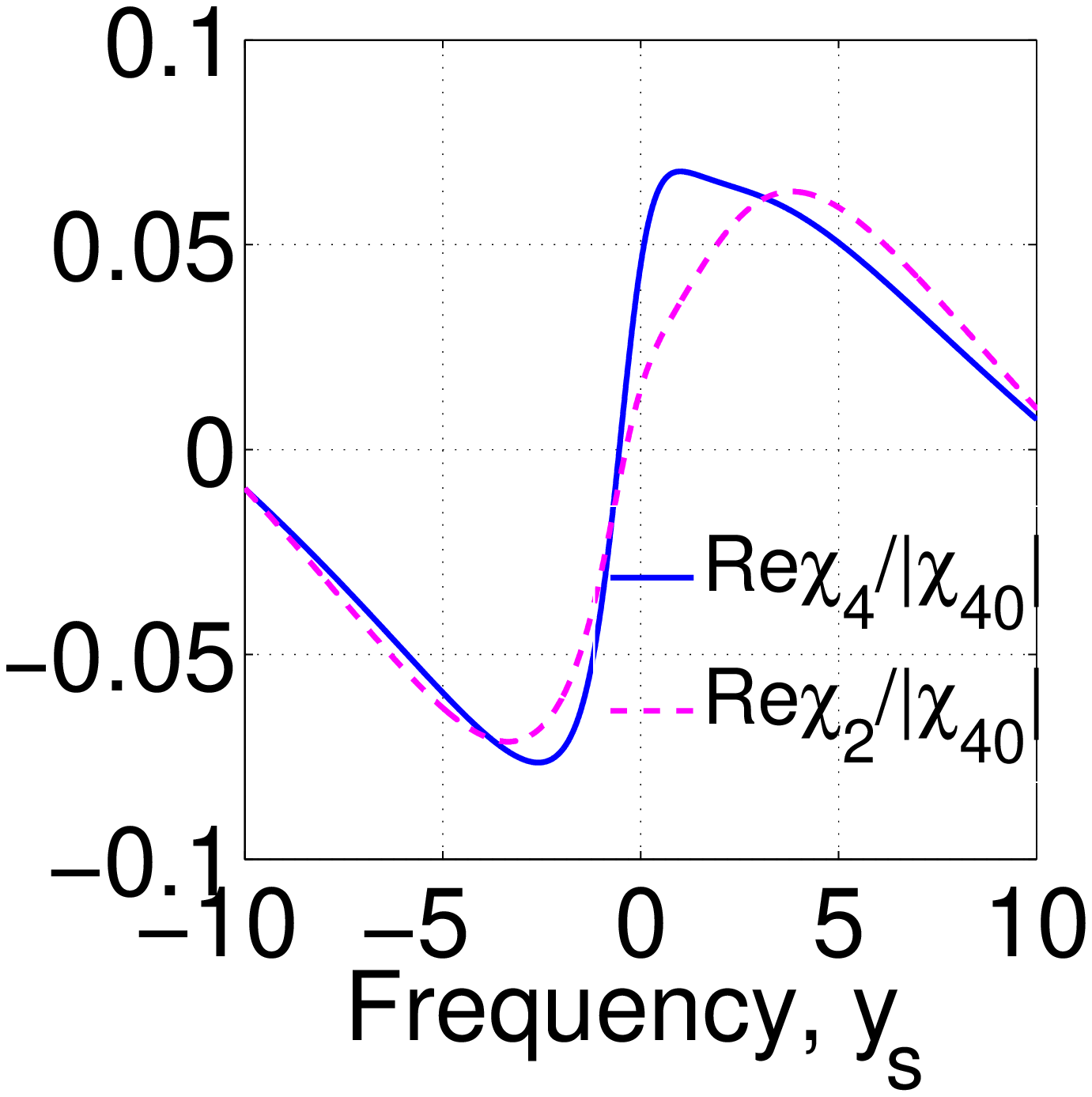}
\includegraphics[width=.325\columnwidth]{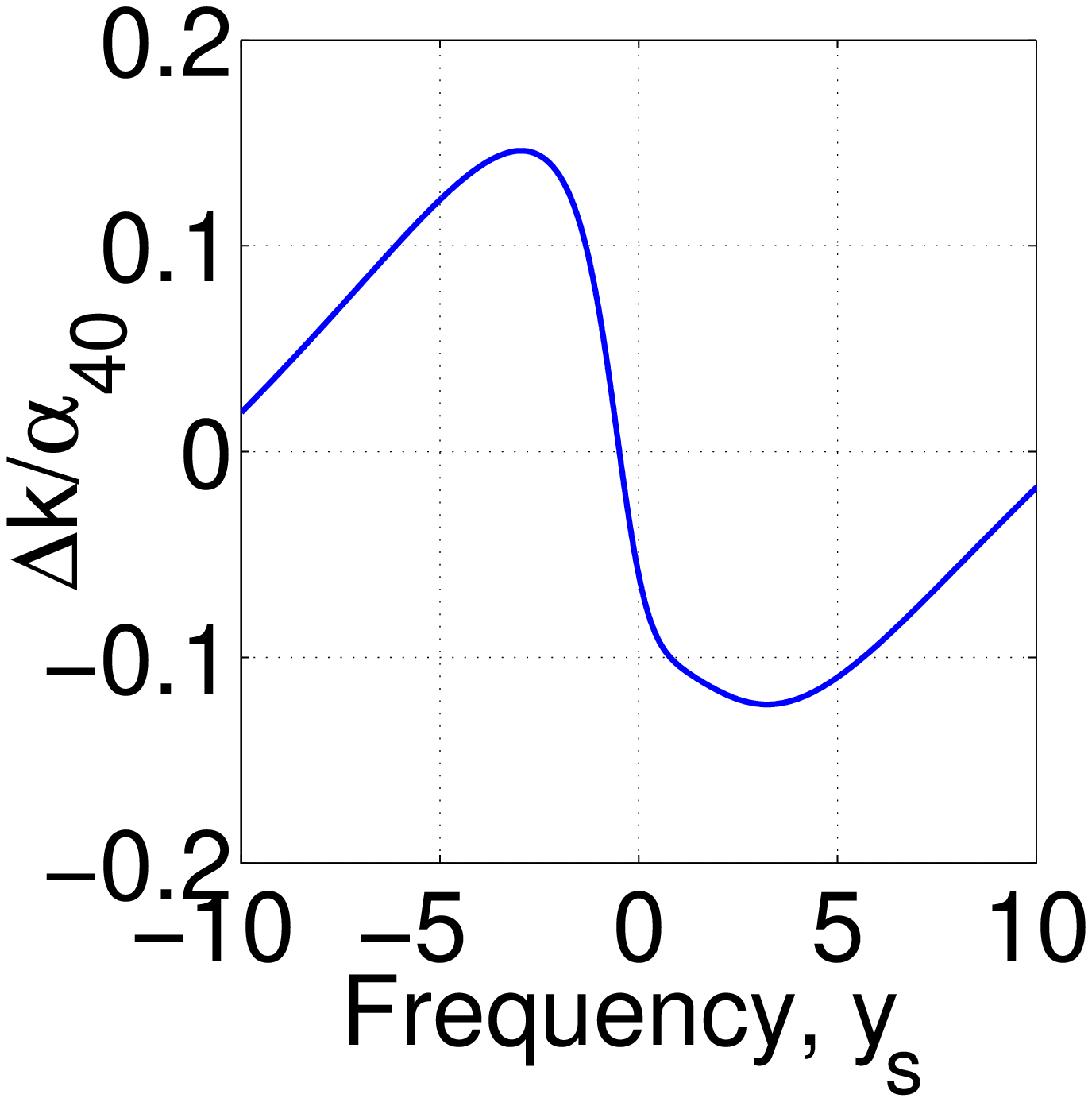}
\\[0pt]
(a) \hspace{20mm} (b)\hspace{20mm} (c)\\[0pt]
\includegraphics[width=.325\columnwidth]{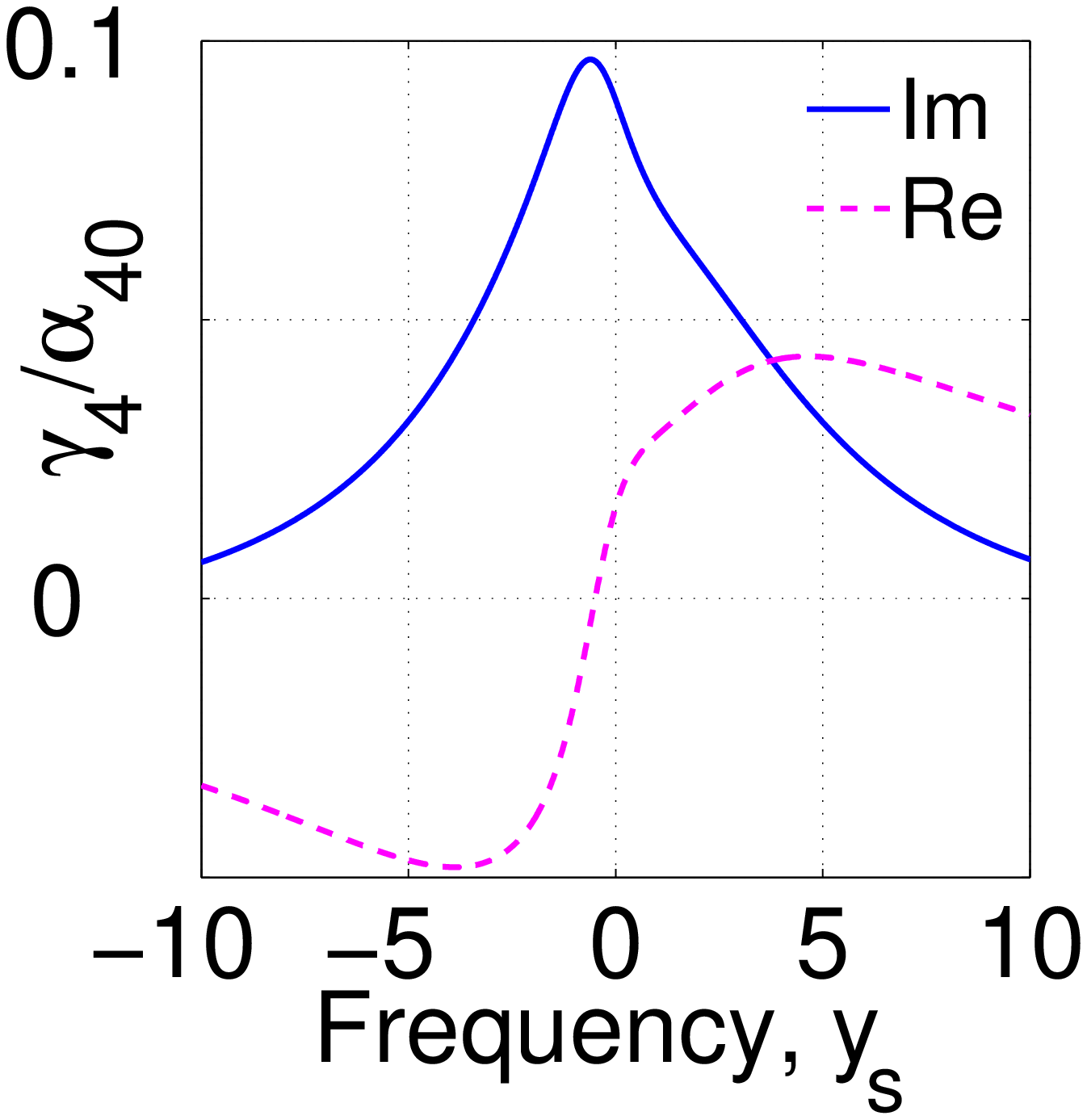}
\includegraphics[width=.325\columnwidth]{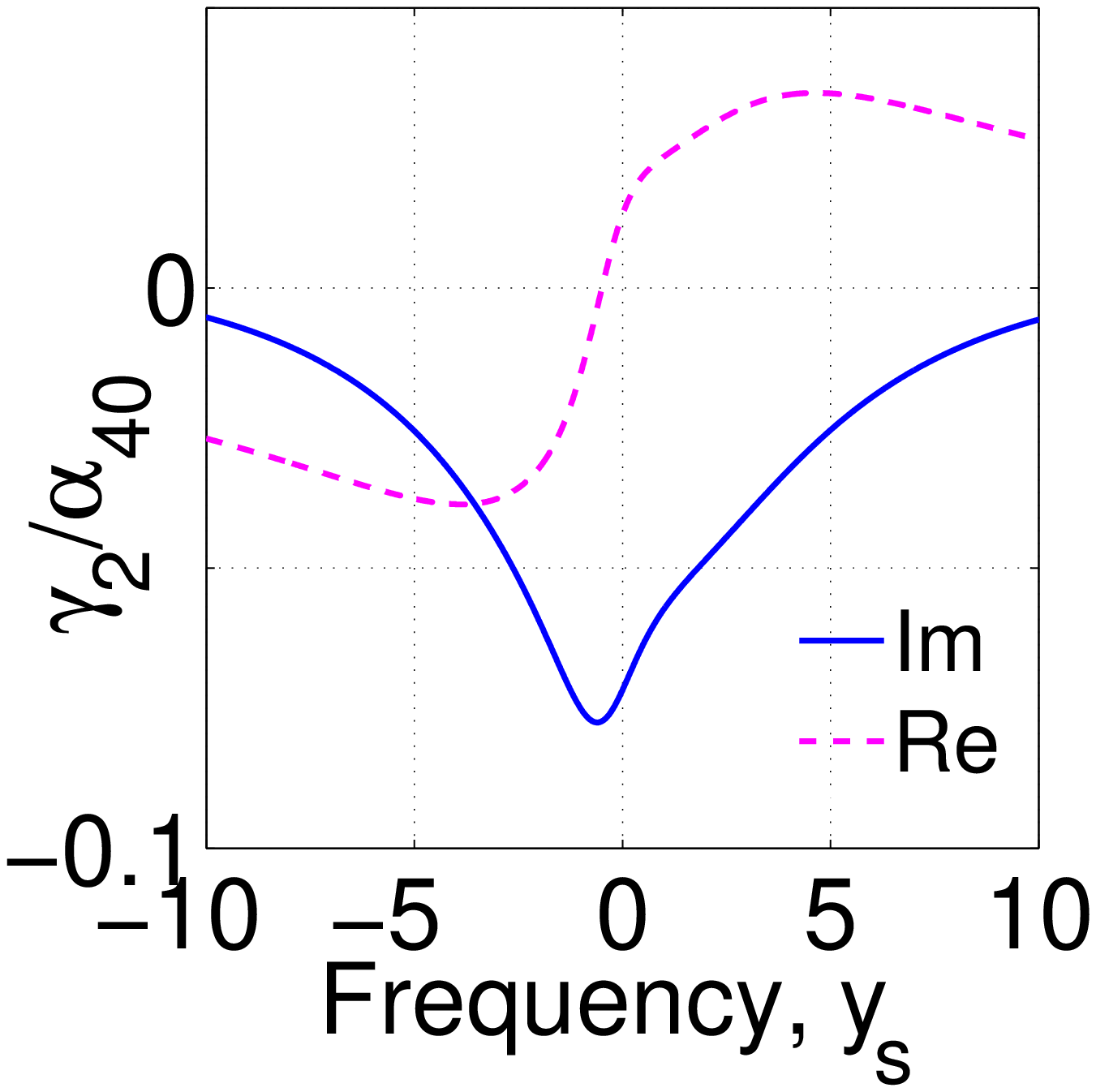}
\includegraphics[width=.325\columnwidth]{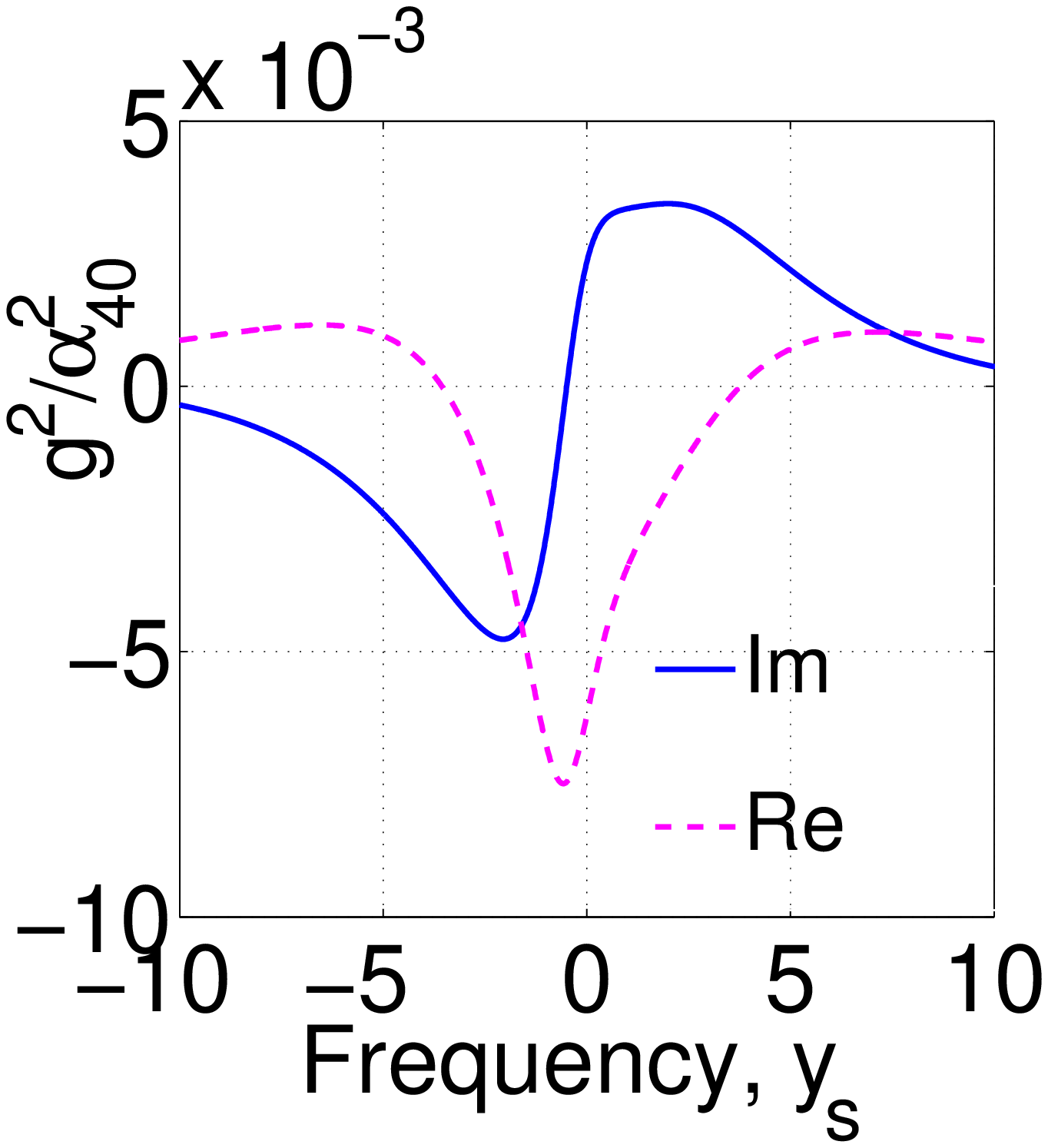}
\\[0pt]
(d) \hspace{20mm} (e)\hspace{20mm} (f)\\[0pt]
\end{center}
\caption{\label{f6} Nonlinear interference structures in absorption of the signal and amplification of the idler, (a); in refractive indices and phase mismatch attributed to the embedded centers, (b) and (c); and in the nonlinear coupling parameters, (d)-(f). $y_s=\Omega_4/\Gamma_{ln}$. $\Omega_1=-\Omega_3=\Gamma_{ln}$, $G_1=21$
~GHz, $G_3=27$~GHz. }
\end{figure}
In this case, nonlinear interference structures with different spectral widths only partially overlap each other. Typical results for the numerical experiments are given below. Population transfer, power broadening of the nonlinear interference structures, and the overall power-saturation of the local optical and NLO parameters depend on both the intensities and resonance frequency offsets of the control fields. This determines different sets for the optimum parameters for coherent energy exchange between the control fields, backward signal and the idler.
\begin{figure}[h!]
\begin{center}
\includegraphics[width=.325\columnwidth]{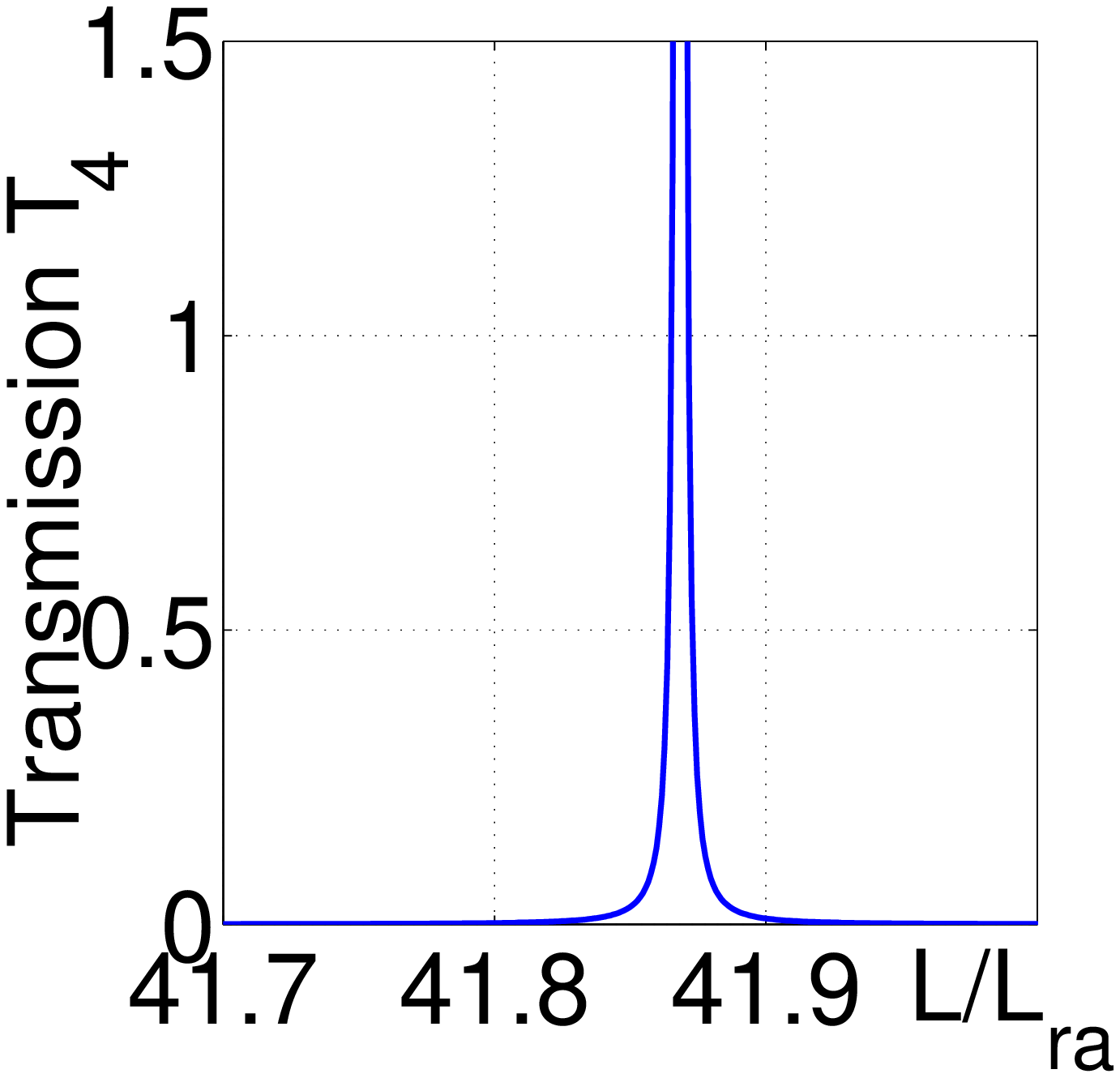}
\includegraphics[width=.325\columnwidth]{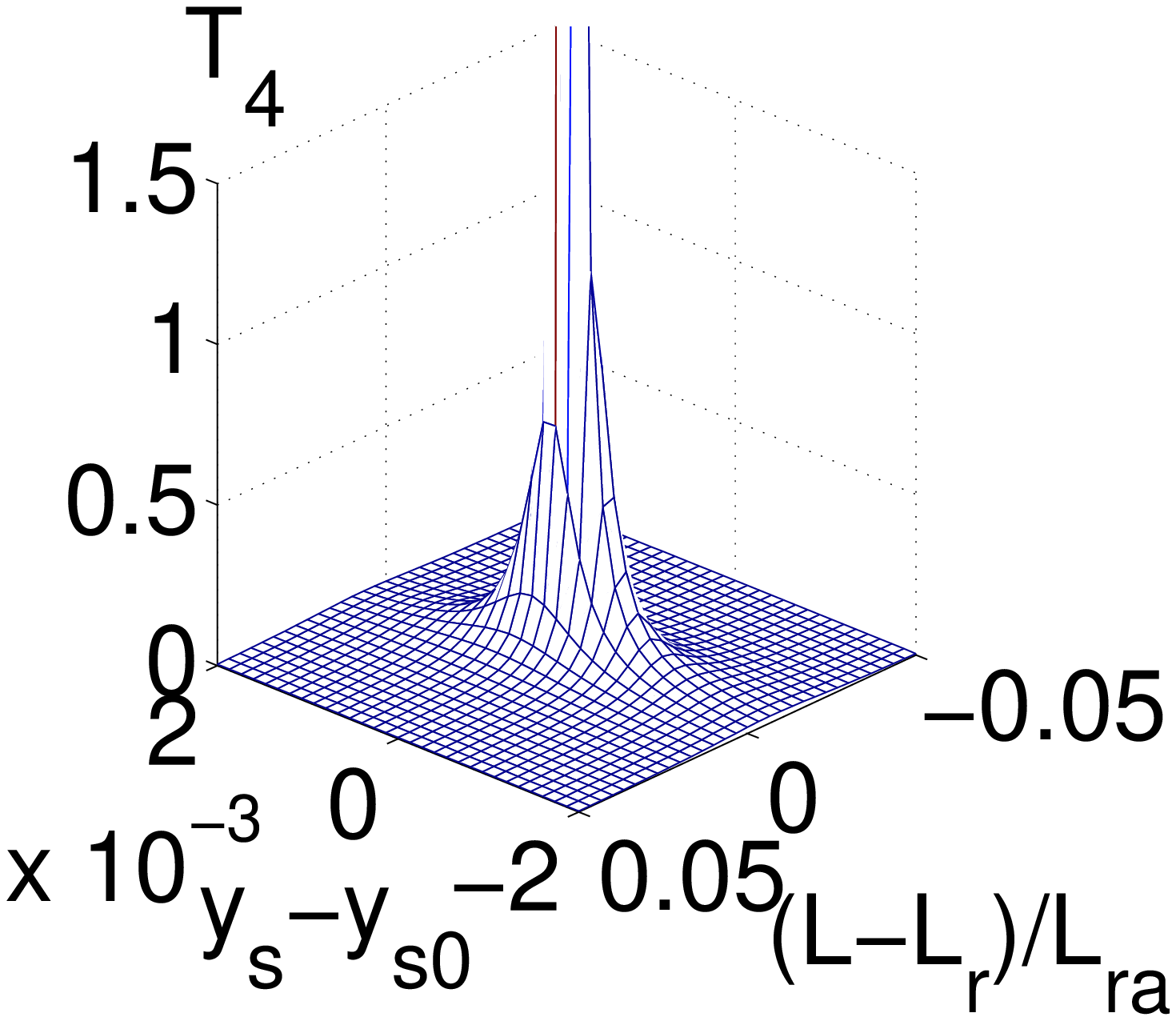}
\includegraphics[width=.325\columnwidth]{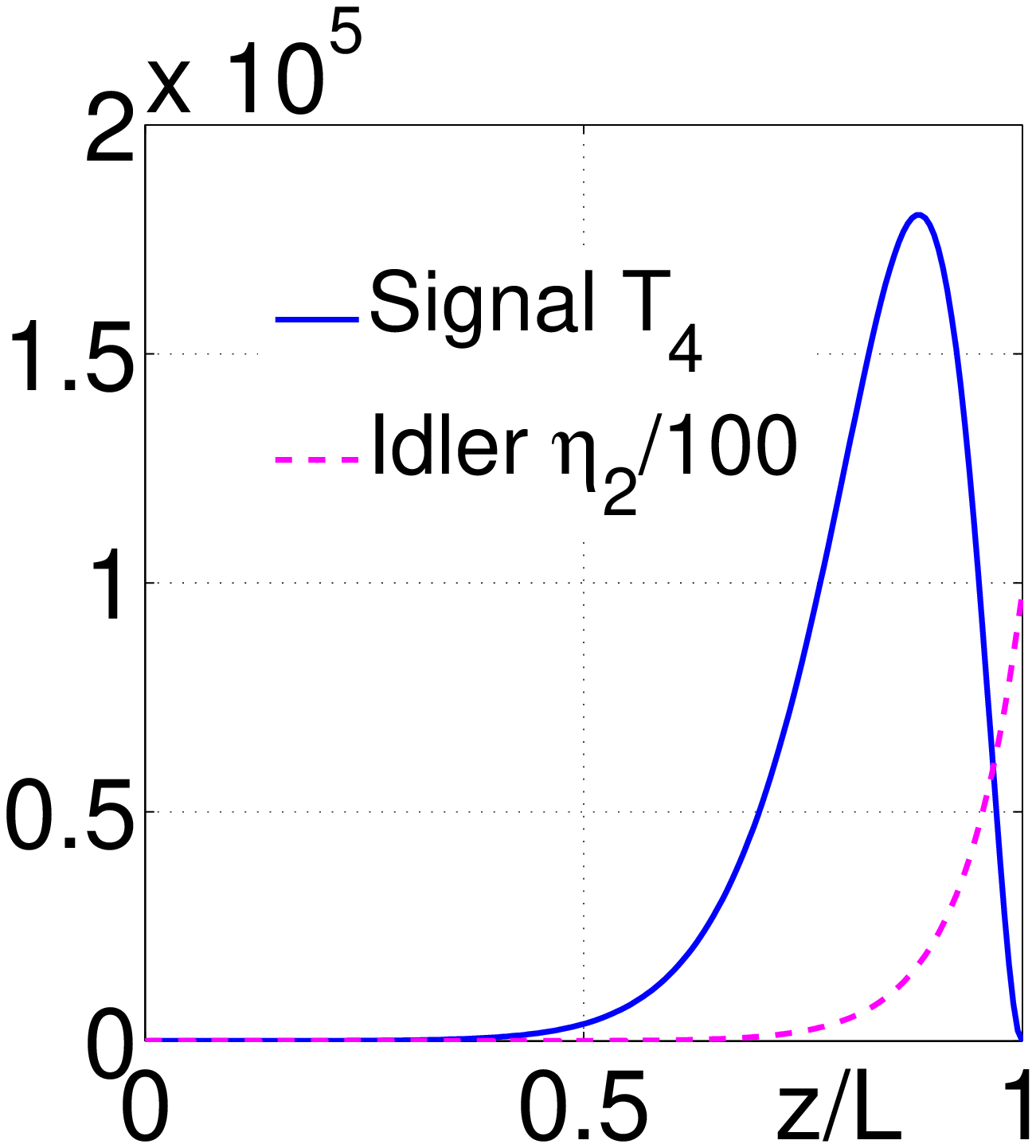}
\\[0pt]
(a) \hspace{20mm} (b)\hspace{20mm} (c)\\[0pt]
\end{center}
\caption{\label{f7} Geometrical and frequency resonances in the output signal and idler, (a) and (b); and distribution of the signal and idler inside the slab at L slightly different from L$_r$, (c). $y_s=\Omega_4/\Gamma_{ln}$. $\Omega_1=-\Omega_3=\Gamma_{ln}$, $G_1=21$~GHz, $G_3=27$~GHz, $\protect\alpha_{NIM}L=2.3$, $\protect\delta k/\protect\alpha_{40}=-0.013069$. (a) and (c):
$y_{s0}=-5.624$. (b): $L_r/L_{ra}=41.86$ (c): $L/L_{ra}=41.865$}
\label{f7}
\end{figure}
Figure~\ref{f6} displays an appreciable change in the local optical parameters at the given frequency detunings and the optimum intensity of the control fields as compared with Fig.~\ref{f4}. Consequently, the propagation features change such that the maximum transparency and amplification of the signal become achievable at a slab optical density about three times smaller $L_r=41.86$~$\alpha_{40}^{-1}$ [Fig.~\ref{f7}(a)]. Other features are similar to those of Fig. ~\ref{f4}.  Energy-level populations at given strength and resonance offsets of the control fields and other local optical parameters at $y_{s}=y_{s0}$ are as follow: $r_l\approx 0.49$,
$r_g\approx 0.48$,
$r_n\approx 0.015$,
$r_m\approx 0.015$,
$\alpha_4/\alpha_{40}\approx0.38$,
$\alpha_2/\alpha_{40}\approx-0.36$,
$\Delta k/\alpha_{40}\approx 0.15$,
$\Im(\gamma_2/\alpha_{40})\approx- 0.046$,
$\Re(\gamma_2/\alpha_{40})\approx -0.037$,
$\Im(\gamma_4/\alpha_{40})\approx 0.059$,
$\Re(\gamma_4/\alpha_{40})\approx-0.047$,
$\Im(g^2/\alpha_{40}^2)\approx -4.26\times 10^{-3}$,
$\Re(g^2/\alpha_{40}^2)\approx -8.4\times 10^{-4}$.

\subsection{Frequency offsets on the order of the optical transition width}
Here, optimum coupling parameters are analyzed for frequency resonance offsets that are several hundred times larger than those in the previous subcase.
\begin{figure}[!h]
\begin{center}
\includegraphics[width=.325\columnwidth]{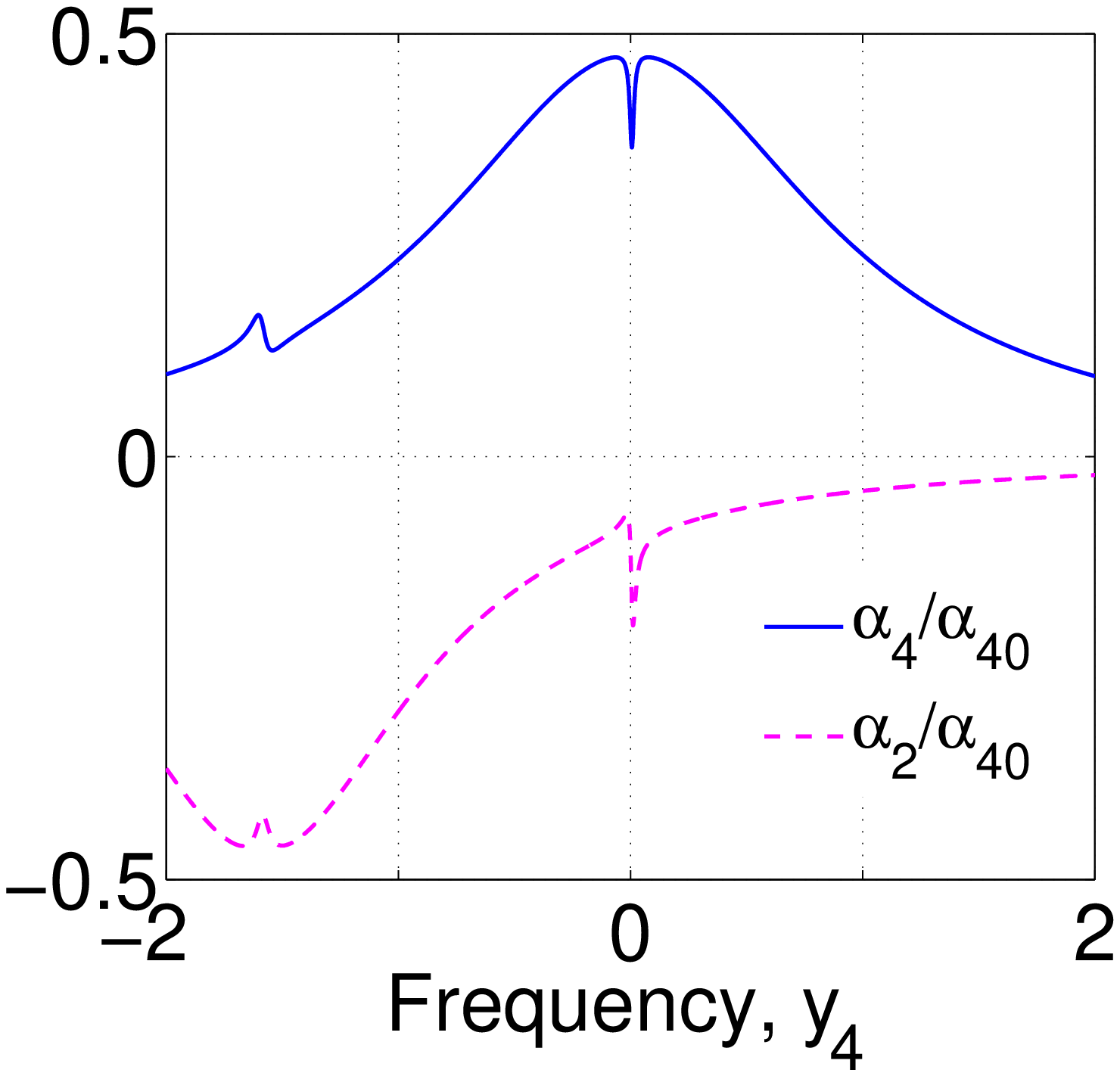}
\includegraphics[width=.325\columnwidth]{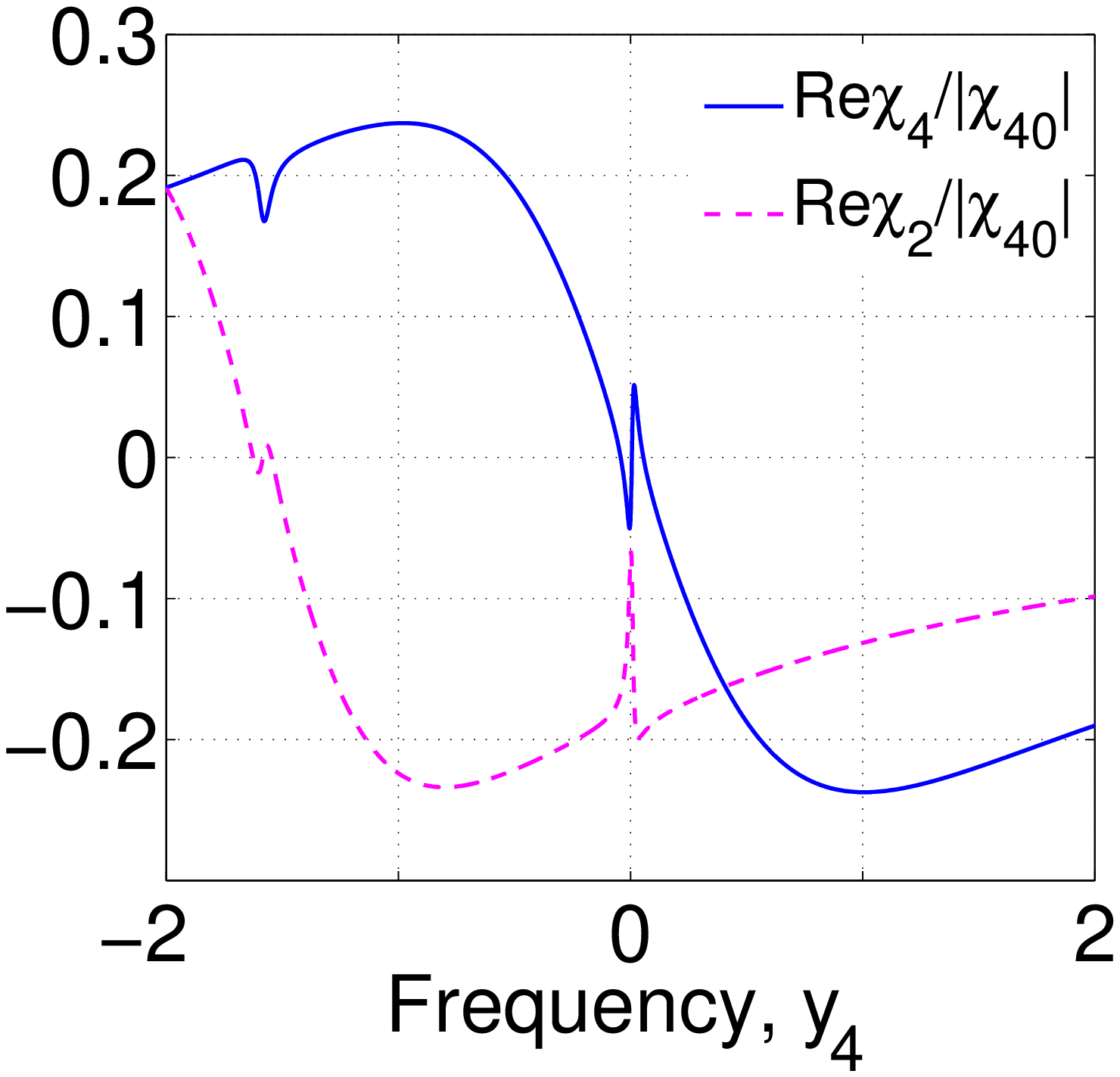}
\includegraphics[width=.325\columnwidth]{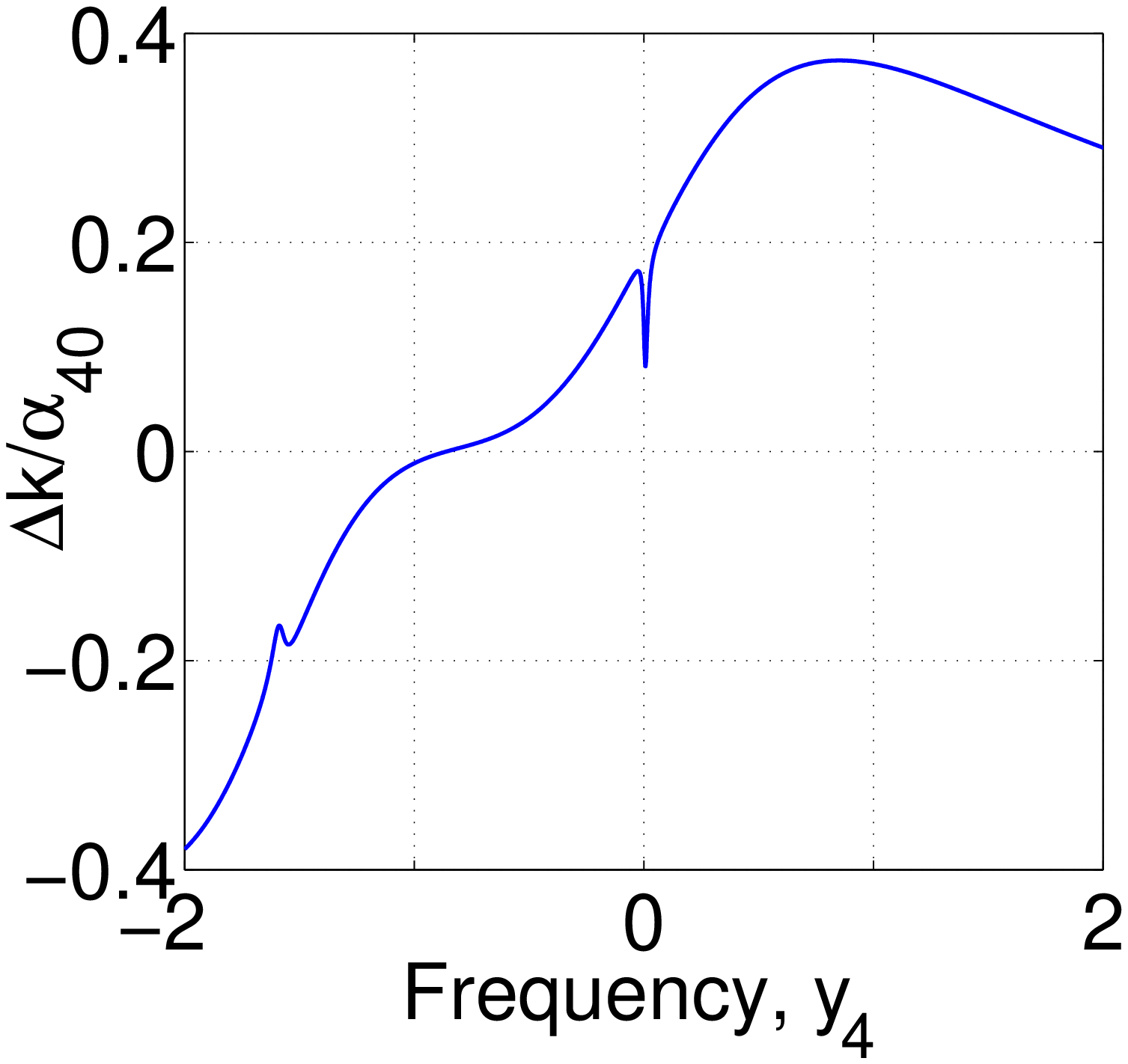}
\\[0pt]
(a) \hspace{20mm} (b)\hspace{20mm} (c)\\[0pt]
\includegraphics[width=.325\columnwidth]{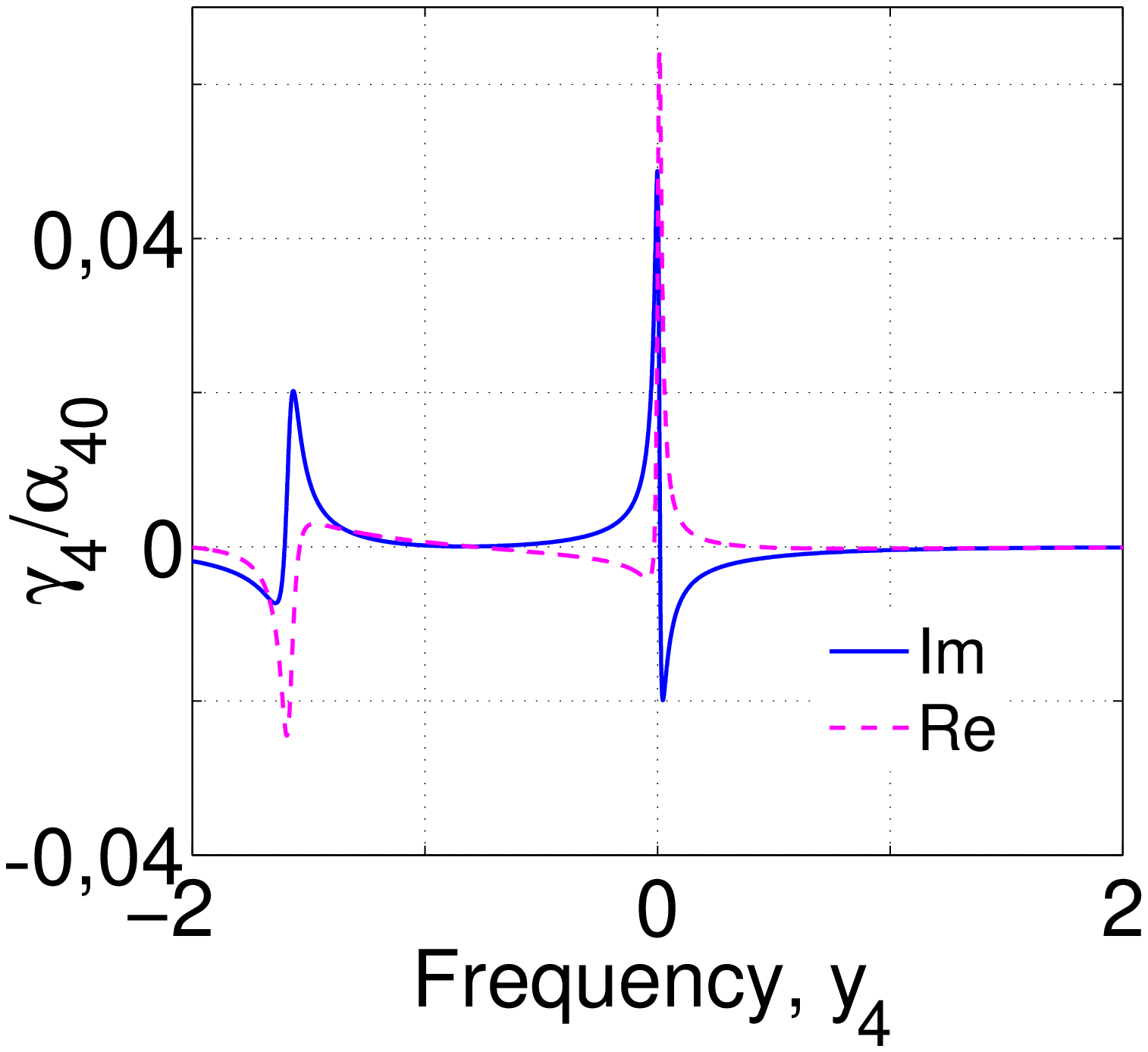}
\includegraphics[width=.325\columnwidth]{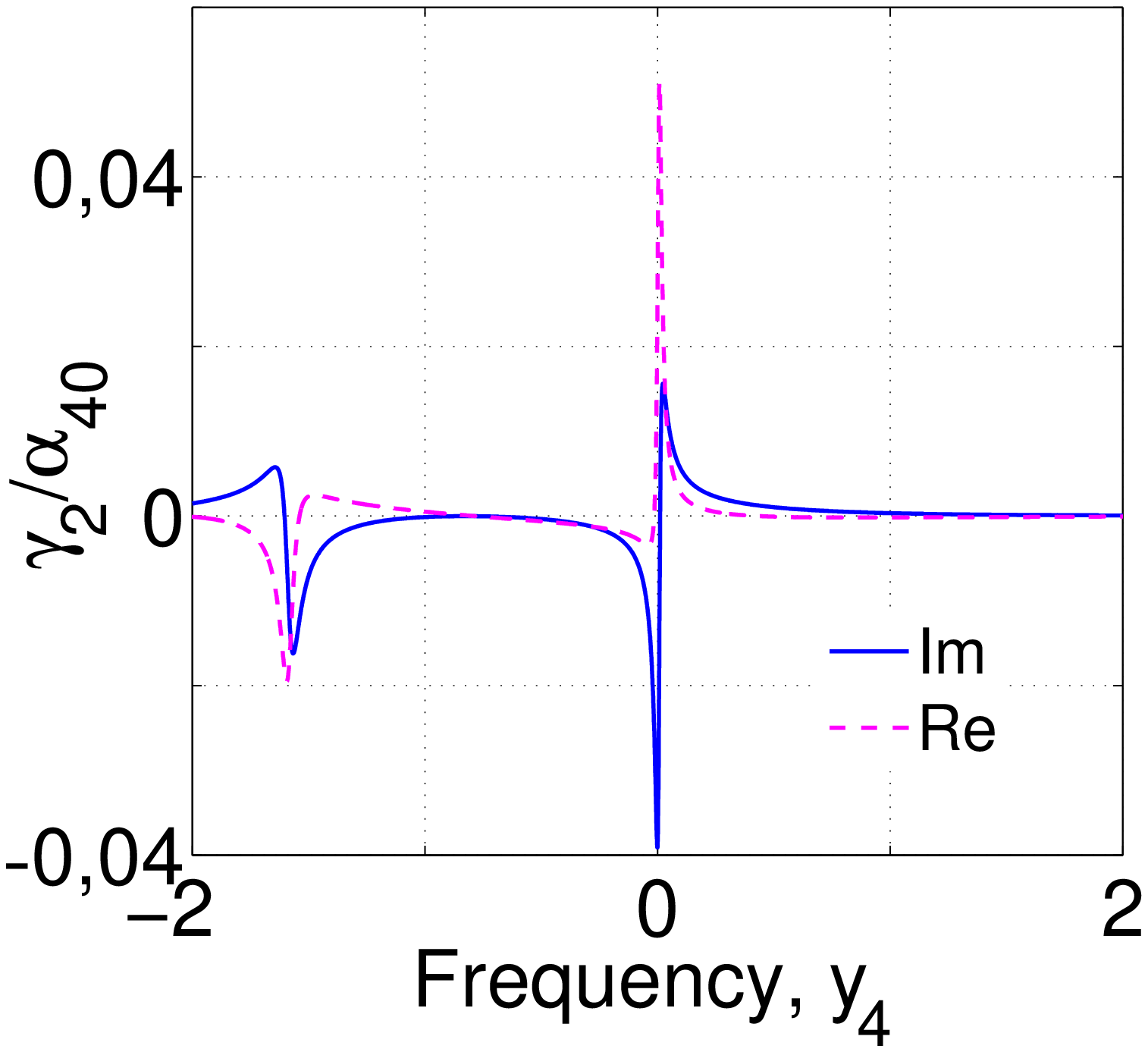}
\includegraphics[width=.325\columnwidth]{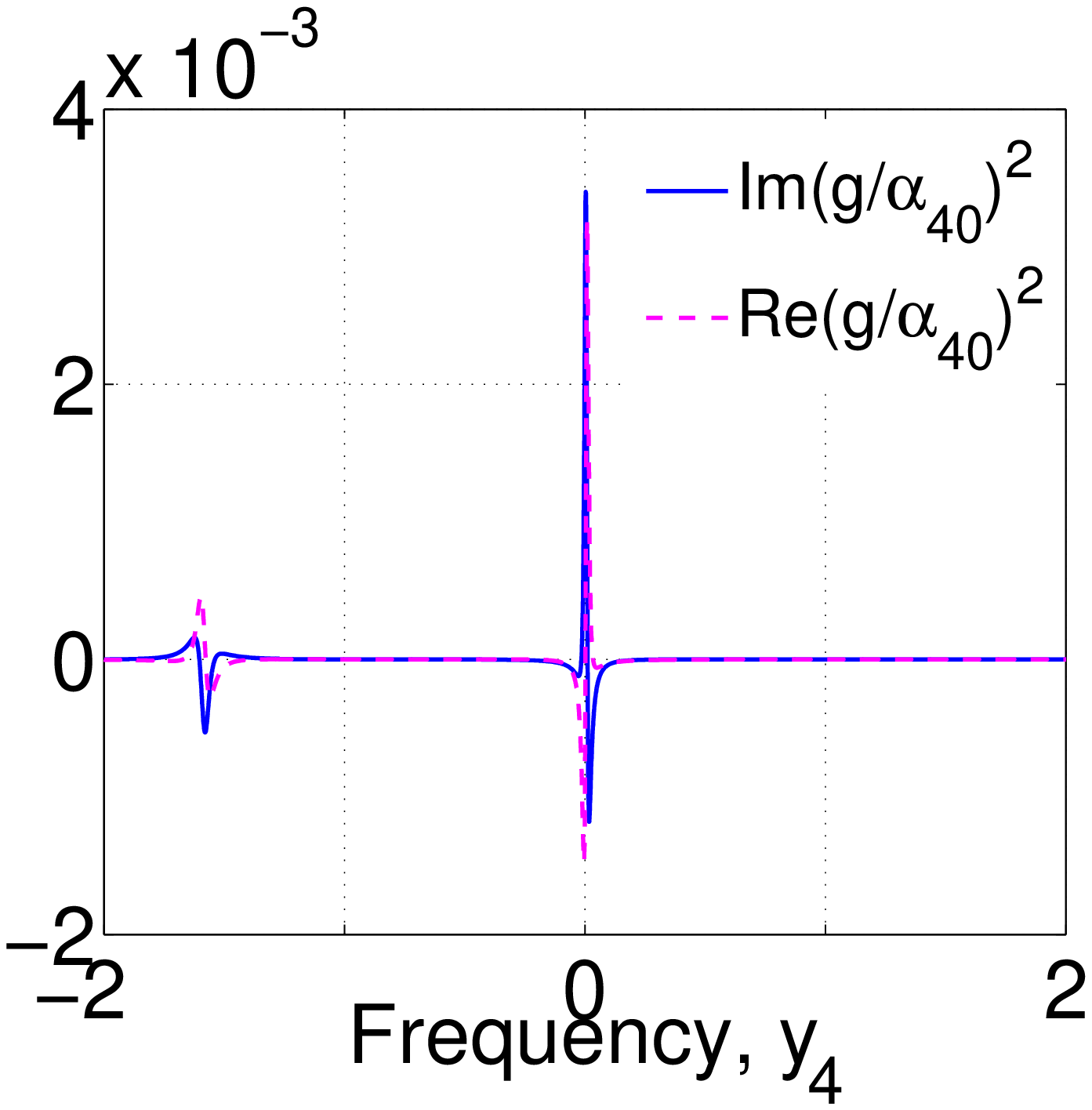}
\\[0pt]
(d) \hspace{20mm}(e) \hspace{20mm} (f)
\end{center}
\caption{\label{f8} Scaled absorption index for the signal and amplification
index for the idler (a); changes in the refraction indices
(b) and phase mismatch (c); NLO coupling parameters, (d) (e);
and parameter $g^2$, (f), in the presence of the control fields at
$G_1=33.19$~GHz, $G_3 = 15$~GHz, $\Omega_1 = -3\Gamma{gl}$, $\Omega_3=0$.
}
\end{figure}
\begin{figure}[!h]
\begin{center}
\includegraphics[width=.325\columnwidth]{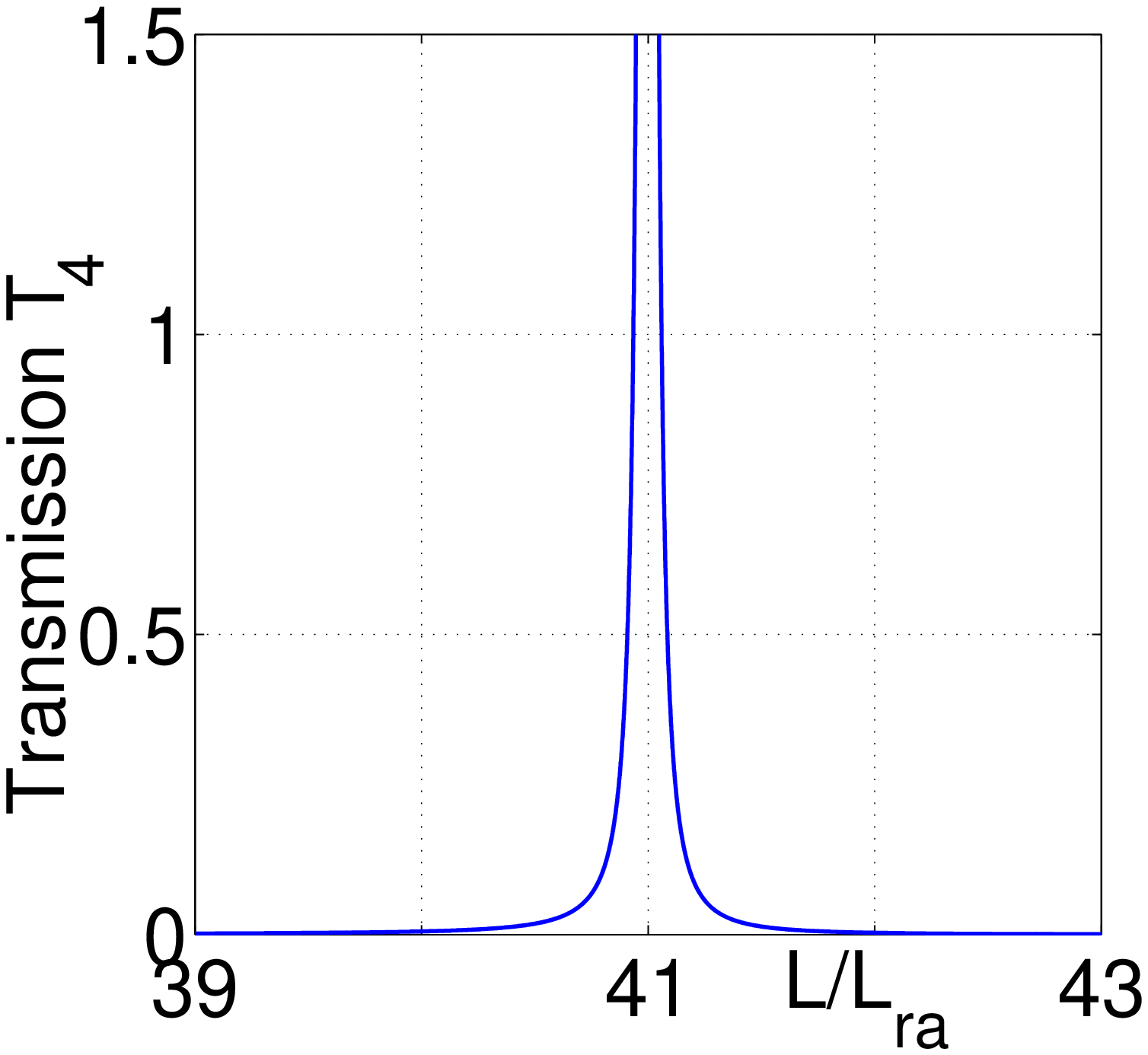}
\includegraphics[width=.325\columnwidth]{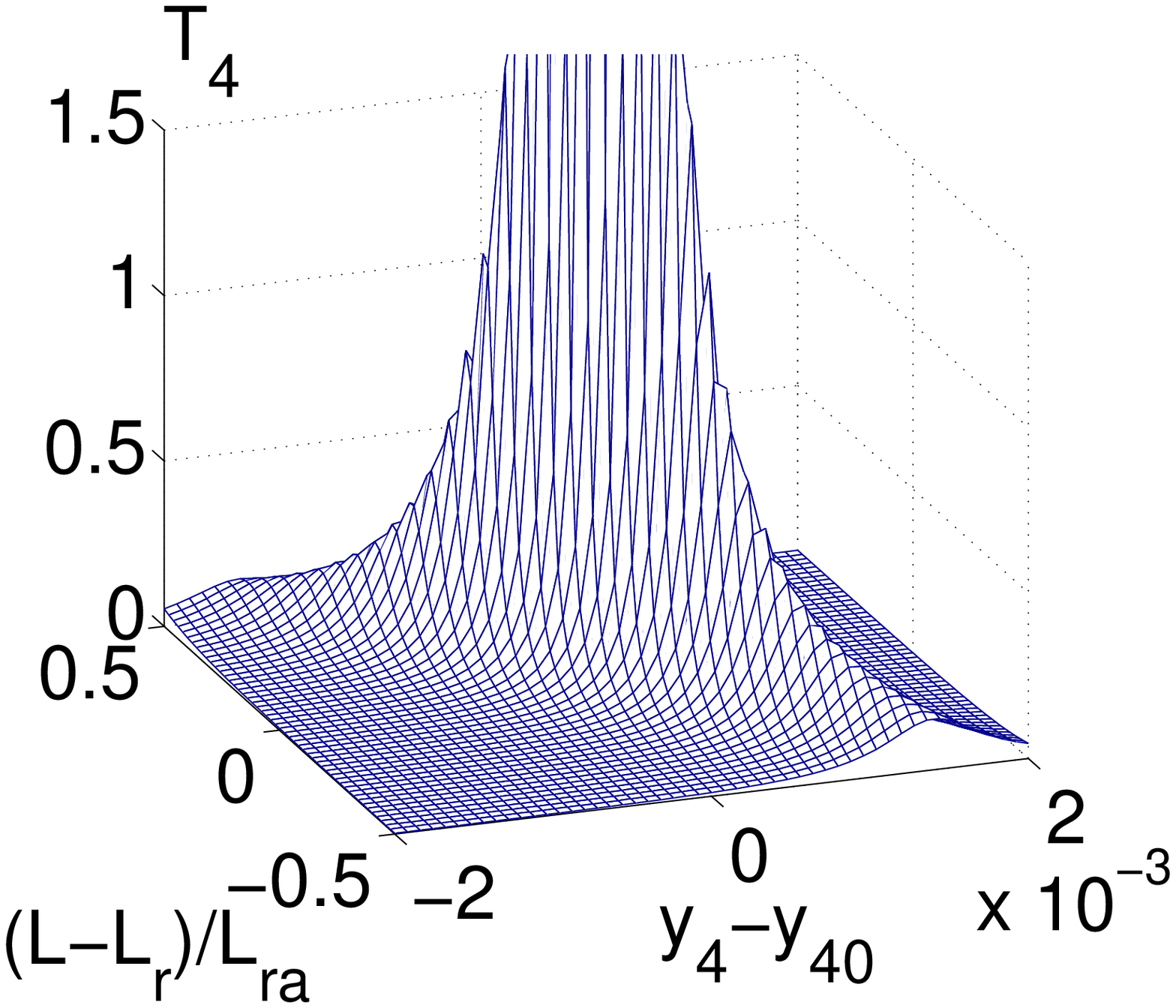}
\includegraphics[width=.325\columnwidth]{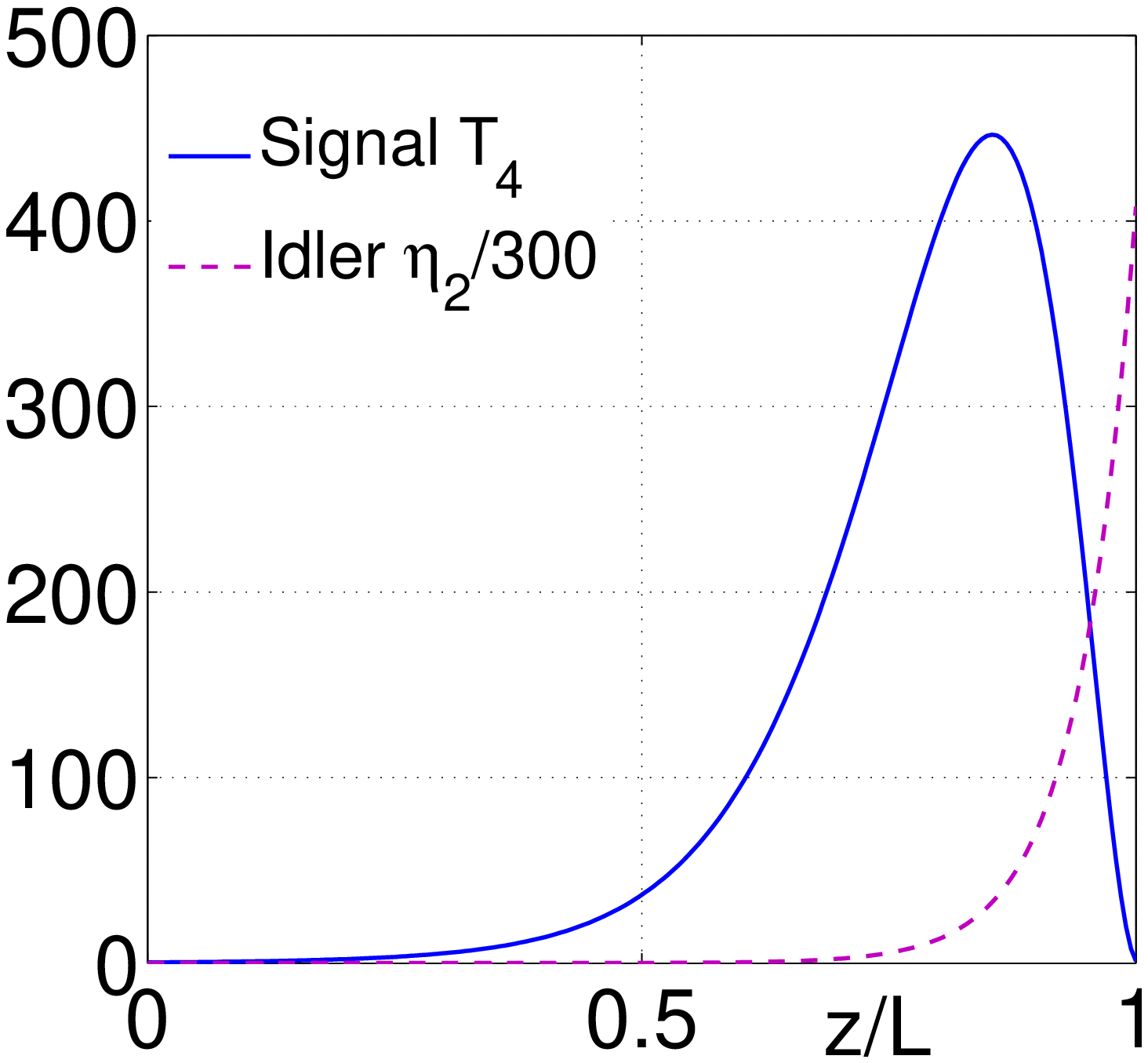}\\
(a) \hspace{20mm}(b)\hspace{20mm}(c)
\end{center}
\caption{\label{f9} Geometrical and frequency resonances in the output signal and idler, (a) and (b); and distribution of the signal and the idler inside the slab at L slightly different from $L_r$, (c). $y_4 =\Omega_4/\Gamma_{ml}$, $\Omega_1 = -3\Gamma{gl}$, $\Omega_3=0$, $G_1=33.19$~GHz, $G_3 = 15$~GHz. $\alpha_{NIM}L =2.3$, $y_{40}=-1.578947$, $\delta k/\alpha_{40} = 3.165\times10^{-4}$. (a) and (c): $y_4 = y_{40}$. (b): $L_r/L_{ra} = 40.9977$. (c): $L/L_{ra} = 40.8977$.}
\end{figure}
Figure~\ref{f8} displays the appearance of two nonlinear resonances and an increase in the amplification of the idler. At the given parameters, the control fields cause essential change in the level populations, $r_l\approx0.49$; $ r_g\approx0.48$; $r_n\approx0.015$; $r_m\approx0.015$, which is accompanied by the large population inversion at the idle transition.
Figure \ref{f9} shows the transmitted signal at the frequency close to the maximum NLO response of the embedded molecules, which may become possible owing to the appropriate phase mismatch, $\delta k$, introduced by the host material. Figure~\ref{f9}(a) displays a narrow geometrical resonance at $ L\approx 41 L_{ra}$ for the optimum frequency offset for the signal at $\Omega_4$=-1.578947~ $ \Gamma_{ml}$. This provides the optimum compensation the given phase mismatch, $\delta k$, introduced by the host material. Changes in $\Delta k$ introduced by molecules are accounted for within the simulations. Figures~\ref{f9}(b) and (c) depict qualitatively similar behavior as in the previous subcases. The calculated optical parameters for $y_4 = y_{40}$ are as follow:
$\Im(\gamma_2/\alpha_{40})\approx-0.014$,
$\Re(\gamma_2/\alpha_{40})\approx-0.015$,
$\Im(\gamma_4/\alpha_{40})\approx 0.017$,
$\Re(\gamma_4/\alpha_{40})\approx-0.019$,
$\Im(g^2/\alpha_{40}^2)\approx-5.31\times 10^{-4}$,
$\Re(g^2/\alpha_{40}^2)=6.25\times 10^{-5}$,
$\alpha_4/\alpha_{40}=0.15$,
$\alpha_2/\alpha_{40}=-0.42$,
$\Delta k/\alpha_{40}=-0.17$.

Amplification in the output signal maxima in Figs.~\ref{f4}, ~\ref{f6} and \ref{f8} reaches many orders of magnitude, which indicates the feasibility of oscillations and, hence, cavity-less generation of counter-propagating signal and idler photons. It is known that even weak amplification per unit length may lead to lasing provided that the corresponding frequency is in a high-quality cavity or feedback resonances, which is equivalent to a great extension of the effective length of a low-amplifying medium. Assuming a resonance absorption cross-section $\sigma_{40}~\sim~10^{-16}$~cm$^2$, which is typical for dye molecules, and a concentration of molecules $ N~\sim~10^{19}$ cm$^{-3}$, we obtain $\alpha_{40}~\sim 10^3$~cm$^{-1}$, and the required slab thickness is in the range $L~\sim (10 - 100) \mu$. The contribution to the index of refraction by the impurities is estimated as $ \Delta n < 0.5(\lambda/4\pi)\alpha_{40}\sim 10^{-3}$, which essentially does not change the negative refractive index.

\section{Conclusion}

In conclusion, we have investigated the feasibility of all-optical manipulation of the optical properties of NIMs through coherent nonlinear-optical energy transfer from the control to the signal field. The strong nonlinear optical response of the composite is primarily determined by the embedded resonant four-level nonlinear centers and, hence, can be adjusted independently. In addition, we have shown the opportunity for quantum control of the local optical parameters, which employs constructive and destructive quantum interference tailored by two auxiliary driving control fields. Such a possibility is proven with the aid of a realistic numerical model. The investigated features are promising for the compensation of losses in strongly absorbing NIMs, which is the key problem that limits numerous revolutionary applications of this novel class of electromagnetic materials. Among the other possible applications are a novel class of the miniature frequency-tunable narrow-band filters, quantum switchers, amplifiers and cavity-free microscopic optical parametric oscillators that allow the generation of entangled counter-propagating left- and right-handed photons. The unique features of the proposed photonic devices are revealed, such as the strongly resonant behavior with respect to the material thickness, the density of the embedded resonant centers and the intensities of the control fields, the feasibility of negating the linear phase-mismatch introduced by the host material, and the important role of the supplementary nonparametric amplification of the idler.

\section*{Acknowledgments}

This work was supported by the U.~S. Army Research Laboratory and by the U.
S. Army Research Office under grants number W911NF-0710261 and 50342-PH-MUR.

\end{document}